\documentclass[reviewcopy]{elsarticle}

\usepackage[reviewcopy]{adndt}
\usepackage{longtable}
\usepackage{comment}
\usepackage{amsmath}
\usepackage{amssymb}
\usepackage{xcolor}
\usepackage{subcaption}
\usepackage{pdflscape}
\usepackage{rotating}
\usepackage{ulem}
\biboptions{square,sort&compress}
\bibpunct[]{[}{]}{,}{n}{}{;}
\begin{document}

\begin{frontmatter}

\journal{Atomic Data and Nuclear Data Tables}


\title{\normalfont\textsc{Tune-out and magic wavelengths, and electric quadrupole transition properties of the singly charged alkaline-earth metal ions}}

\author{Mandeep Kaur$^a$}
\author{Sukhjit Singh$^a$}
\author{B. K. Sahoo$^b$}
\author{Bindiya Arora$^a$}

\ead{E-mail: bindiya.phy@gndu.ac.in}
\address{$^{a}$Guru Nanak Dev University, Amritsar, Punjab-143005, India.}
\address{$^{b}$Atomic, Molecular and Optical Physics Division, Physical Research Laboratory, Navrangpura, Ahmedabad-380009, India.}
\cortext[cor1]{Guru Nanak Dev University, Amritsar, Punjab-143005}

\date{today} 

\begin{abstract}
In continuation to our earlier reported data on the electric dipole (E1) matrix elements and lifetimes of the metastable states of the alkaline earth 
ions in [Atomic Data and Nuc. Data Tables {\bf 137} (2021) 101381], we present here the tune-out and magic wavelengths of the Mg$^+$, Ca$^+$, Sr$^+$ 
and Ba$^+$ alkaline earth-metal ions by determining dynamic E1 polarizabilities. Furthermore, we have evaluated the electric quadrupole (E2) matrix 
elements of a large number of forbidden transitions using an all-order relativistic many-body method and compare them with the previously reported 
values for a few selective transitions. Compilation of both the E1 and E2 transition matrix elements, will now provide a more complete knowledge
about the transition properties of the considered singly charged alkaline earth-metal ions. Similarly, the listed precise values of tune-out and magic
wavelengths due to the dominant E1 polarizabilities can be helpful to conduct experiments using the above ions with reduced systematics. Therefore,
all these data will be immensely useful for various applications for carrying out the high-precision experiments and laboratory simulations in atomic 
physics, and interpreting transition lines in the astrophysical observations.
\end{abstract}


\end{frontmatter}

\tableofcontents
\listofDtables
\listofDfigures
\vskip5pc

\section{Introduction}

Accurate knowledge of spectroscopic properties like transition probabilities, oscillator strengths, static and dynamic polarizabilities, etc. of 
atomic systems, particularly of the singly charged alkaline-earth metal ions, possess immense importance as these are the potential candidates 
for conducting high-precision fundamental experiments. Owing to the innovation in laser cooling and trapping techniques, these ions have created 
benchmark in the field of quantum state manipulation experiments \cite{nigg, wilson}, atomic clocks~\cite{chou}, investigating nuclear charge radii 
\cite{ruiz2016, shi2017}, studying parity \cite{fortson} and Lorentz-invariance symmetry violations \cite{wood,tiecke} and so on. The systematic 
fractional uncertainties of $10^{-18}$ are expected to be attained in the frequency standards based on the frequency measurements of the forbidden 
transitions of the cold singly charged Ca$^+$ \cite{huang,champenois} and Sr$^+$ \cite{barwood,fordell,dube} ions. The accurate estimates of the 
systematics due to Stark effects, quadrupole shifts and black-body radiation (BBR) shifts are essential to reach the asserted uncertainty in these 
ions. Estimation of the Stark and BBR shifts require dipole polarizabilities, whereas estimation of quadrupole shifts require quadrupole moments and 
quadrupole polarizabilities. These quantities can be determined with the help of a large number of electric dipole and quadrupole transition matrix 
elements.

The static dipole polarizability values for $3S_{1/2}$, $3P_{3/2,1/2}$ and $3D_{5/2,3/2}$ states of Mg$^+$ were calculated by Mitroy et al.
\cite{mit2009} using a semi-empirical single electron analysis combined with the relativistic all-order single-double method (MBPT-SD). The 
theoretical calculations for the static dipole polarizabilities of Ca$^+$ for the $4S_{1/2}$, $4P_{3/2,1/2}$ and $3D_{5/2,3/2}$ states were given by 
Tang et al.~\cite{tang13}, Mitroy et al.~\cite{mitroy2008} and Sahoo et al.~\cite{sahoo2009a} using relativistic structure model, diagonalizing 
semi-empirical Hamiltonian and relativistic coupled-cluster (RCC) methods respectively. The spectral analysis technique measurement for the static 
dipole polarizability for the ground state of Ca$^+$ is given by Chang et al.~\cite{chang83}. Similarly, for Sr$^+$, RCC calculations by Sahoo et
al.~\cite{sahoo2009b}, relativistic all-order calculations by Jiang et al.~\cite{dansha2009} and semi-empirical calculations by Mitroy et 
al.~\cite{mitroy2008PRA} are available for the $5S$ and $4D$ states. Also, Safronova and co-workers~\cite{safr2010} have reported dipole 
polarizabilities for the $(6-8)S$, $(5-7)P$ and $(4-6)D$ states for Sr$^+$. Experimental results by Barklem et al. and Nuankaew et al. 
\cite{barklem,nunkaew2009} are also available. Sahoo et al.~\cite{sahoo2009b} and Barrett et al.~\cite{barrett2019} have calculated polarizabilities for the $6S$ and $5D$ states of 
Ba$^+$. Moreover, the dipole polarizability of the $6S$ state of Ba$^+$ is also reported by Safronova et al. using relativistic all-order method~\cite{isk2008} whereas, Stark ionization spectroscopy measurement by Snow et al.~\cite{snow2007} and radio frequency resonance measurement by Gallagher et al.~\cite{gallagher1982} for the $6S$ state of Ba$^+$ are also given. 
Static quadrupole
polarizability for the ground state of Mg$^+$ have been calculated by Mitroy et al.~\cite{mitroy2008} using the semi-empirical Hamiltonian method and 
by Sahoo et al.~\cite{sahoo2007} using the RCC method. Relativistic all-order calculations by Safronova et al.~\cite{safronova2011} and semi-empirical
Hamiltonian calculations by Mitroy et al.~\cite{mitroy2008} provided quadrupole static polarizabilities values for the $4S_{1/2}$ state of Ca$^+$. 
Scalar components of the quadrupole polarizability for the $5S$, $5P$ and $4D$ states of Sr$^+$ were reported by Jiang et al~\cite{jiang2016}. The 
RCC calculations by Sahoo et al.~\cite{sahoo2012}, relativistic all-order method calculations by Iskrenova et al.~\cite{isk2008} and experimental 
measurements by Snow et al.~\cite{snow2007} are available for the $6S_{1/2}$ ground state of Ba$^+$.

At the tune-out wavelengths ($\lambda_{\rm{T}}$), the dynamic dipole polarizability of atomic systems vanishes. Similarly, the differential dynamic 
polarizabilities nullify at the magic wavelengths ($\lambda_{\rm{magic}}$). Both these quantities provide state-insensitive trapping to reduce the 
large systematics in high-precision experiments. Theoretical relativistic structure model calculations by Tang et al.~\cite{tang13} and relativistic 
configuration interaction plus core-polarization (RCICP) calculations by Jiang et al.~\cite{jiang2017} as well as experimental measurements by Liu et 
al.~\cite{liu2015} have given these values for Ca$^+$, whereas, Jiang et al.~\cite{jiang2016} have predicted $\lambda_{\rm{magic}}$ values for Sr$^+$.
Moreover, the tune-out wavelengths $\lambda_{\rm{T}}$ are also presented for some of the states of considered alkaline earth metal ions for linearly polarized light in the Ref.~\cite{jasthesis}. Apprehension of these $\lambda_{\rm{T}}$ values are prerequisite for the sympathetic cooling of 
possible singly and multiply charged ions in two-species mixtures~\cite{rosenband,hobein}. 

In the earlier theoretical studies, Stark shifts of the clock transitions in the alkaline-earth metal ions were determined considering contributions 
only from the ionic dipole polarizabilities. But, the next generation of high-precision optical frequency standards today face the challenge of 
systematics due to the stray electric field gradients to which atoms or ions are susceptible. In an optical lattice ion clock, quadrupole shifts due 
to the stray electric field and gradients produced from the neighboring ions can significantly affect the determination of stark shifts. Even though 
the contributions from the quadrupole polarizabilities are orders of magnitudes smaller than that of the dipole polarizabilities, they do have 
non-negligible contributions to the total Stark shifts. There are two most important reasons for which contributions from quadrupole polarizability 
are often neglected in the estimations of the Stark shifts. First, these contributions were assumed to be negligible at the previously considered 
precision level of interest. Second, it is cumbersome to determine the tensor components of quadrupole polarizability of atomic states. But, it is possible to determine these quantities now by extending our recent derivations on evaluating higher-components of static quadrupole polarizabilities \cite{sukhjit2018}.

Owing to the revolution in the observational techniques, it has become possible to observe weak and forbidden quadrupole transitions from astronomical 
objects~\cite{wood1916,borsenberger,mashonkina1999}. Thus, the forbidden transition properties are of great astrophysical interests. In interstellar medium, many alkaline earth metals are present in the abundance predominantly in the singly ionized form that is Mg$^+$~\cite{wood1916}, 
Ca$^+$~\cite{mashonkina2008,murga2015}, Sr$^+$~\cite{andrievsky,mashonkina2001} and Ba$^+$~\cite{mashonkina2008} and are considered as the significant
constituent of interstellar medium. Information about the structure and physical characteristics of these interstellar clouds can be inferred using
the quadrupole transition probabilities of these ions~\cite{welty,kwan2011}. Forbidden emission lines of Mg$^+$ from the metastable levels are 
important as they are abundant in the Solar flares~\cite{sandlin1976}. Electric quadrupole transition probabilities are important in the plasma 
diagnostic studies, plasma temperature and dynamics~\cite{edlen1984}. Owing to the long coherence time of the ground and metastable states, the 
quadrupole transitions of the alkaline-earth metal ions are used for encoding the quantum bit in order to realize quantum logic techniques
\cite{roos2006,benhelm2008}. Calculations of electric quadrupole transitions provide better insights of the roles of electron correlation effects in 
the atomic systems~\cite{sahoo2006PRA,sahoo2017}. Since atomic clock frequency measurements are aimed at achieving below $10^{-19}$ precision level today, it is absolutely necessary to investigate effects like contributions from the applied electric field-gradients in the experiments that would appear at such unprecedented precision level. These quadrupole energy shifts can be evaluated
using quadrupole polarizabilities. Long-range interatomic interactions, ion mobility, van der Waals constant between different systems, scattering 
properties of collisions between a neutral atom and its ions~\cite{cote} can be evaluated using quadrupole polarizability. To realize all these 
properties, quadrupole matrix elements are needed. However, only a few quadrupole matrix elements for ground and metastable states of the Ca$^+$, 
Sr$^+$ and Ba$^+$ ions are available. 

A few groups have studied the quadrupole transition probabilities for Mg$^+$, Ca$^+$, Sr$^+$ and Ba$^+$ alkaline-earth metal ions using relativistic 
and non-relativistic methods. The weakest bound electron potential model theory (WBEMPT) calculations of Celik et al.~\cite{celik}, the RCC 
calculations by Majumder et al.~\cite{sanjoy} and non-relativistic multi-configuration Hartree-Fock calculations by Fischer et al.~\cite{fischer} have 
provided quadrupole transition probabilities of the Mg$^+$. The semi-empirical core-potential calculations for Ca$^+$, Sr$^+$ and Ba$^+$ are provided 
by Filippin et al.~\cite{filippin} for the ground state to metastable state transition probabilities. Similarly, the RCC calculations by Guan et 
al.~\cite{guan2015} for Ca$^+$ and pseudo-relativistic Hartree-Fock method calculations by Gurell et al.~\cite{gurell2007} for Ba$^+$ are given for 
the transition probabilities between ground state and metastable states. 
 
In this work, we intend to present data tables for the accurate values of quadrupole matrix elements for a large number of transitions of the above 
ions, E2 radiative properties, tune-out wavelengths and magic wavelengths of the Mg$^+$, Ca$^+$, Sr$^+$ and Ba$^+$ ions. We have estimated the quadrupole matrix elements and
other related radiative properties of the transitions involving the $nS_{1/2}-n'D_{5/2,3/2}$, $n'D-n''D$, $n'D_{5/2,3/2}-mG_{9/2,7/2}$, 
$qP-q'P$ and $qP-m'F$ E2 transitions with $\{n,n',n''\}= 3-6$, $m=5,6$, $\{q,q'\}= 3-5$ and $m'=4-5$ in Mg$^+$; $n=4-7$, $\{n',n''\}= 3-6$, 
$m=5,6$, $\{q,q'\}= 4-6$ and $m'=4-5$ in Ca$^+$; $n=5-8$, $\{n',n''\}= 4-7$, $m=5,6$, $\{q,q'\}= 5-8$ and $m'=4-5$ in Sr$^+$; and $n=6-9$, 
$\{n',n''\}= 5-8$, $m=5,6$ and $\{q,q'\}= 6-8$ in Ba$^+$. The static dipole polarizabilities for the $nS_{1/2}$, $n'P_{3/2,1/2}$ and 
$mD_{5/2,3/2}$ states with $n=3-5$, $n'=3-5$ and $m=3-4$ in Mg$^+$; $n=4-6$, $n'=4-6$ and $m=3-5$ in Ca$^+$; $n=5-8$, $n'=5-7$ and $m=4-6$ in Sr$^+$ 
and $n=6-8$, $n'=6-8$ and $m=5$ in Ba$^+$. The static quadrupole polarizabilities are given for the  $nS$, 
$n'P$ and $mD$ states with $\{n,n',m\}=3$ in Mg$^+$; $\{n,n'\}=4$ and $m=3$ in Ca$^+$; $\{n,n'\}=5$ and $m=4$ in Sr$^+$ and $n=6$ and $m=5$ in Ba$^+$.
The dynamic dipole polarizabilities for the transitions $nS-nP$ and $nS-nD$ in case of Mg$^+$ (n=3), whereas for the transitions $nS-nP$ and 
$nS-(n-1)D$ in case of Ca$^+$ (n=4); Sr$^+$ (n=5) and Ba$^+$ (n=6) are presented using the dipole matrix elements given in our 
previous work \cite{mandeep}. Using these dipole dynamic polarizabilities, the magic wavelengths are obtained by locating the crossings between the 
dynamic polarizabilities of the $nS_{1/2}$, $nP_{1/2}$, $nP_{3/2}$, $nD_{3/2}$ and $nD_{5/2}$ states of Mg$^+$(n=3); $nS_{1/2}$, 
$nP_{1/2}$, $nP_{3/2}$, $(n-1)D_{3/2}$ and $(n-1)D_{5/2}$ states of Ca$^+$(n=4), Sr$^+$(n=5) and Ba$^+$(n=6) plotted against the wavelengths in the 
range 300-1300 nm. The dipole dynamic polarizabilities are also plotted to evaluate tune-out wavelengths for the $nS$, $n'P$ and $mD$ 
states having $\{n,n',m\}=3$ in Mg$^+$; $\{n,n'\}=4$ and $m=3$ in Ca$^+$; $\{n,n'\}=5$ and $m=4$ in Sr$^+$ and $\{n,n'\}=6$ and $m=5$ in the case of 
Ba$^+$. The dynamic quadrupole polarizabilities are also presented for the Ca$^+$, Sr$^+$ and Ba$^+$ ions, which involve transitions between 
the ground state and metastable states. 

\section{Theory and Methods for calculations} \label{sec2}

In this section, we mention the procedures adopted for calculating atomic wave functions, which are then used to evaluate the electric dipole 
(E1) and quadrupole (E2) matrix elements for different transitions. Thereafter, general formulae for the calculation of transition probabilities, 
oscillator strengths and static as well as dynamic dipole and quadrupole polarizabilities are given. All the quantities are listed in atomic 
units (a.u.) unless otherwise mentioned.

\subsection{Relativistic all-order method}

The atomic wave functions, which are prerequisite for the evaluation of all the above mentioned spectroscopic properties, are evaluated by employing 
the relativistic all-order method~\cite{Blundell,theory, PhysRevA.91.042507,PhysRevA.92.052511}. We have considered only the singles and doubles 
excitations through this approach (SD method), in which the wave function of an atomic state having a closed-shell configuration and a valence orbital
is given by 
\begin{eqnarray}
|\Psi_v \rangle_{\rm SD} &=& \left[1+ \sum_{ma}\rho_{ma} a^\dag_m a_a+ \frac{1}{2} \sum_{mnab} \rho_{mnab}a_m^\dag a_n^\dag a_b a_a + 
\sum_{m \ne v} \rho_{mv} a^\dag_m a_v + \sum_{mna}\rho_{mnva} a_m^\dag a_n^\dag a_a a_v\right] |\Phi_v\rangle,
 \label{expansion}
\end{eqnarray}
where $|\Phi_v\rangle$ is the reference state function and defined as $|\Phi_v\rangle = a_v^{\dag}|0_c\rangle$ with the Dirac-Hartree-Fock (DHF) 
wave function of the closed-core of the respective ion $|0_c\rangle$, $a^\dag_i$ and $a_i$ are the creation and annihilation second quantization 
operators, respectively, with subscripts $m,n,\cdots$ and $a,b,\cdots$ refer to the virtual and occupied orbitals, respectively, of the reference 
state, and the index $v$ represents for the valence orbital. In the above expression, $\rho_{ma}$ and $\rho_{mv}$ are amplitudes of the single 
excitations involving core and valence electrons, respectively. Similarly, $\rho_{mnab}$ and $\rho_{mnva}$ are the amplitudes of the double 
excitations involving core and core with valence electrons. These excitation coefficients are obtained by solving the Schr\"odinger equation for the 
Dirac-Coulomb Hamiltonian in the iterative procedure till self-consistent solutions are achieved.

In order to verify contributions from the next level of excitations in the all-order method, we have added important triple excitations in the 
perturbative approach (SDpT method) be defining additional term as defined below
\begin{equation}
|\Psi_v\rangle_{\rm SDpT}\approx |\Psi_v\rangle_{\rm SD}+\left[ \frac{1}{18}\sum_{mnrabc}\rho_{mnrabc}a_m^{\dag}a_n^{\dag}a_r^{\dag}a_ca_ba_a    
+ \frac{1}{6}\sum_{mnrab}\rho_{mnrvab}a_m^{\dag}a_n^{\dag}a_r^{\dag}a_ba_aa_v  \right] |\Phi_v\rangle,
\label{expansion1}
\end{equation}
where $\rho_{mnrabc}$ and $\rho_{mnrvab}$ denote the amplitudes of the perturbative triple excitations involving the core and core with valence 
electrons, respectively.

We have used 70 B-splines of order $k = 11$ for each angular momentum in order to obtain the single particle orbitals in the DHF method. We have 
defined the radial functions on a non-linear grid, which are constrained to large spherical cavity having radius $R = 220$ a.u.. A sufficiently 
large number of virtuals are accommodated by this choice of cavity radius and quality of the orbitals are ensured due to consideration of a 
sufficiently large number of B-splines.
   
\subsection{Evaluation of matrix elements}\label{matrix}

After obtaining the wave functions either from the SD or SDpT approximations, we evaluate transition matrix element of an one-body operator $O$ 
between the states $|\Psi_v\rangle$ and $|\Psi_w\rangle$ as
\begin{equation}
 O_{vw} = \frac{\langle \Psi_v|O|\Psi_w\rangle}{\sqrt{\langle\Psi_v|\Psi_v\rangle\langle\Psi_w|\Psi_w\rangle}},
 \label{wavef1}
\end{equation}
where $O$ corresponds to either of the E1 or E2 operators. After substituting expressions from Eqs. (\ref{expansion}) and (\ref{expansion1}) 
for the SD and SDpT expressions, one can classify these contributions into core and valence correlation contributions. Thus, it gives rise to 
two terms have dominant contributions to the transition matrix elements and those are \cite{CC}
\begin{eqnarray}
 O^{(a)}_{vw}=\sum_{ma}(o_{am} \tilde{\rho}_{vmwa}+o_{ma}{\tilde{\rho}^*}_{wmva})
 \end{eqnarray}
 and
 \begin{eqnarray}
 O^{(c)}_{vw}=\sum_{m}(o_{vm} \rho_{mw}+o_{mw}\rho^*_{mv}),
\end{eqnarray}
where $\tilde{\rho}_{vmwa}= \rho_{vmwa}-\rho_{mvaw}$ and $*$ represents the complex conjugate term. In our approach, $o_{vw}$ gives rise the lowest 
order DHF values to the estimated matrix elements.

To estimate contributions from the neglected higher-level excitations, we have also carried out semi-empirical scaling of the wave functions in the 
SD and SDpT methods. For this purpose, we only modify the dominant contributions to single valence excitations by defining as \cite{Dansha2008} 
\begin{eqnarray}
\rho_{mv}'=\rho_{mv}\frac{\delta E_v^{\rm{expt}}}{\delta E_v^{\rm{theory}}} ,
\end{eqnarray}
where $\rho_{mv}'$ are the modified single excitation coefficients that are used to recalculate the matrix elements and the modified values are 
referred to as the ``scaled" matrix elements. When the scaled wave functions from the SD and SDpT methods are used for the evaluation of matrix 
elements, the corresponding results are termed as $O^{\rm{SD}}_{\rm{sc}}$ and $O^{\rm{SDpT}}_{\rm{sc}}$, respectively. The recommended values for 
the matrix elements are given by comparing the ratio $R=O_{vw}^{(c)}/O_{vw}^{(a)}$. If $R > 1$, then the $\rm SD_{sc}$ values are regarded as the
final ($O^{\rm{Final}}$) values, otherwise the SD results are used as the $O^{\rm{Final}}$ values. This procedure can be found in more detail 
elsewhere (e.g. see Refs. \cite{CC, Blundell}).

\subsection{Electric quadrupole transition properties}

The E2 transition probability ($A^{E2}_{vk}$) in inverse second (s$^{-1}$) from an upper energy level described by state wave function $|\Psi_v\rangle$
with angular momentum $J_v$ to a lower energy level with wave function $|\Psi_k \rangle$ and angular momentum $J_k$, in terms of the fundamental 
constants, is given by \cite{kelleher}
\begin{equation}
A^{E2}_{vk}=\frac{1}{120} \alpha c \pi \sigma \times \left (\frac{\alpha \sigma}{R_{\infty}} \right)^{4} \times \frac{S^{E2}}{g_v} ,
\end{equation}
where $\alpha  = \frac{e^{2}}{4 \pi \epsilon_{0} \hbar c}$ is the fine structure 
constant, $R_{\infty} = \frac{\alpha^{2}m_{e} c}{2h}$ is the Rydberg constant, $c$ is the speed of light and $\sigma=E_v-E_k$ is the energy difference between the upper ($E_v$) and lower ($E_k$) levels of the 
transition and $S^{E2}=|\langle J_v||{\bf Q}|| J_k \rangle|^2$ is the line strength with E2 operator ${\bf Q} \equiv \sum_j {\bf q}_j=
-\frac{e}{2} \sum_j (3z^2_j-r_j^2)$. We have used the values of the fundamental constants as $\alpha  = 7.297352\times10^{-3}$, 
$c=29979245800$ cm s$^{-1}$ and $R_{\infty} =1.0973731\times 10^5$ cm$^{-1}$ from Ref.~\cite{mohr} for the evaluations of the E2 transition 
probabilities. By substituting the values of fundamental constants and wavelength ($\lambda$) of the transition in \AA , $A^{E2}_{vk}$ is determined 
using the following formula given in ~\cite{sobelman1979atomic} as 
\begin{equation}
A^{E2}_{vk} =\frac{1.1199 \times 10^{18}}{g_v\lambda^5} \times S^{E2},  \label{e2coeff}
\end{equation} 
where $g_v$ is the degeneracy factor of corresponding state.

Using the above transition probability equation, the absorption oscillator strengths $f_{kv}^{E2}$ for the E2 transition operator are calculated as~\cite{sobelman1979atomic, kelleher} 
\begin{eqnarray}
f_{kv}^{E2} &=& \left(\frac{R_{\infty}}{2c \alpha^{3} \pi} \right)\frac{g_v}{g_k} \times \frac{A^{E2}_{vk}}{\sigma^{2}} \nonumber \\
   &=& 1.4992 \times10^{-16}\times 
\frac{g_v}{g_k} A^{O}_{vk} \lambda^{2}, \label{eqsf}
\end{eqnarray}
which follows that
\begin{eqnarray}
f^{E2}_{kv}&=&\frac{1}{240\alpha}\left(\frac{\alpha \sigma}{R_{\infty}}\right)^{3} \times\frac{S^{E2}}{g_k} \nonumber \\
&=& \frac{167.90}{g_k \lambda^3} \times S^{E2} . \label{eq-osq}
\end{eqnarray}

\subsection{Electric polarizabilities}
 
The application of oscillating electric field of trapping laser beam induces the Stark shift in the energy levels of an atomic or ionic 
system. This Stark shift of an energy level $E_n$ can be quantified as~\cite{bonin,manakov,beloy}
\begin{eqnarray}
\Delta E_n=-\frac{1}{4}\alpha^d_{n}(\omega){\cal E}^2-\frac{1}{16}\alpha^{q}_{n}(\omega){\nabla{\cal E}}^2+ \cdots , \label{eqshift}
\end{eqnarray}
where $\alpha^d_{n}(\omega)$ is the dynamic dipole polarizability, $\alpha^{q}_{n}(\omega)$ is the dynamic quadrupole polarizability, 
${\cal E}$ is the applied electric field strength and \textbf{$\nabla{\cal E}$} is a tensor describing the gradient of the electric field at the
position of the atomic system. For an uniform electric field, the second term on the right side of Eq. (\ref{eqshift}) diminishes as the gradient for 
uniform field is zero. However, for the inhomogeneous electric field, the second term contributes to the differential shift in energy.

\subsubsection{Dipole polarizability}

For an atomic system in state $|\Psi_n\rangle \equiv |\gamma _n J_n M_{J_n}\rangle$ with angular momentum $J_n$ and its azimuthal quantum number 
$M_{J_n}$, the Stark shift due to the electric field of the applied AC electric field at frequency $\omega$ is given by the first term of 
Eq. (\ref{eqshift}). On applying the sum-over-states approach of perturbative theory, the $\alpha^{d}_{n}(\omega)$ can be determined as 
\begin{eqnarray}
\alpha^d_{n}(\omega)&=&-\sum_{m\neq n}{(p^*)_{nm}(p)_{mn}} \times \left[\frac{1}{\delta E_{nm}+\omega}+\frac{1}{\delta E_{nm}-\omega}\right].\label{eq-dip}
\end{eqnarray}
With $(p)_{mn}=\langle \Psi _m|D|\Psi _n\rangle$ as the E1 matrix element {having ${\bf D}=\sum_j {\bf d}_j 
= - e \sum_j {\bf r}_j$ being the electric dipole (E1) operator and  $\delta E_{nm} = E_m^0-E_n^0$ considered as the difference in the unperturbed energies of the 
corresponding states involved in the transition in accordance with the dipole selection rules. For linearly polarized light, we can express $\alpha_{n}^{d}(\omega)$ by rewriting it as~\cite{sukhjitJPB}
\begin{equation}
\alpha^{d}_{n}(\omega)=\alpha_{n}^{d(0)}(\omega)+\frac{3M_{J_n}^2-J_n(J_n+1)}{J_n(2J_n-1)}\alpha_{n}^{d(2)}(\omega),\label{eq14}
\end{equation}
where $\alpha_{n}^{d(0)}(\omega)$ and $\alpha_{n}^{d(2)}(\omega)$ represent the scalar and tensor of rank 2 part of dipole dynamic polarizability and 
can be written as
\begin{equation}
\alpha_{n}^{d(0)}(\omega)= \sum_{m \ne n} W_{n}^{(0)}  \left [\frac{ |\langle \gamma _nJ_n||{\bf D}||\gamma _m J_m \rangle|^2}
{E_n -E_m +\omega}+\frac{ |\langle \gamma _n J_n||{\bf D}||\gamma _m J_m \rangle|^2}{E_n-E_m-\omega}\right] \label{eqpolz1}
\end{equation}
and
\begin{equation}
\alpha_{n}^{d(2)}(\omega)= \sum_{m \ne n} W_{n,m}^{(2)}  \left [\frac{ |\langle \gamma _n J_n||{\bf D}||\gamma _m J_m \rangle|^2}
{E_n -E_m +\omega}+\frac{ |\langle \gamma _n J_n||{\bf D}||\gamma _m J_m \rangle|^2}{E_n-E_m-\omega}\right] \label{eqpolz3}
\end{equation}
with the coefficients
$W_{n}^{(0)} =-\frac{1}{3(2J_n+1)}$ and} $W_{n,m}^{(2)} =2\sqrt{\frac{5J_n(2J_n-1)}{6(J_n+1)(2J_n+3)(2J_n+1)}} \times (-1)^{J_n+J_m+1}
                                  \left\{ \begin{array}{ccc}
                                            J_n& 2 & J_n\\
                                            1 & J_m &1 
                                           \end{array}\right\}.$
The selection rules of the six-j symbol ensures that only the scalar term contributes towards the total dipole polarizability to the 
states with angular momentum $J=1/2$, whereas both the scalar and tensor components contribute to the states with angular momenta 
$J=3/2$ and $J=5/2$. According to the Eq.~(\ref{eq14}), the total polarizability of the $J=3/2$ state is given by $\alpha^{d}_{n}(\omega)
=\alpha^{d(0)}_{n}(\omega)-\alpha^{d(2)}_{n}(\omega)$ for $M_J=\pm 1/2$ and it corresponds to $\alpha^{d}_{n}(\omega)=\alpha^{d(0)}_{n}
(\omega)+\alpha^{d(2)}_{n}(\omega)$  for $M_J = \pm 3/2$. Similarly for the states with $J=5/2$, the total polarizabilities are 
given by $\alpha^{d}_{n}(\omega)=\alpha^{d(0)}_{n}(\omega)-0.8 \alpha^{d(2)}_{n}(\omega)$ for $M_J=\pm 1/2$, $\alpha^{d}_{n}(\omega)=
\alpha^{d(0)}_{n}(\omega)-0.2 \alpha^{d(2)}_{n}(\omega)$ for $M_J=\pm 3/2$ and $\alpha^{d}_{n}(\omega)=\alpha^{d(0)}_{n}(\omega)+
\alpha^{d(2)}_{n}(\omega)$ for $M_J=\pm 5/2$. It can be noted that the static dipole polarizability values correspond to $\omega = 0$.

 For finding the tune-out wavelengths, we first plot the dynamic polarizability of considered states and then identify the value of $\omega$ where 
the polarizability value turns out to be zero. This value  of $\omega$ corresponds to the $\lambda_{\rm{T}}$. Similarly, $\lambda_{\rm{magic}}$ 
corresponds to the null differential dynamic polarizability of a transition at a given value of $\omega$. The differential AC Stark shift of the 
transition involving the ground and an excited state is due to the dominant dipole polarizability contribution is given by 
\begin{eqnarray}\nonumber
\delta (\Delta E)_{ge}^{d} (\omega ) &=&-\frac{1}{4}\left[\alpha_{g}^{d}(\omega)-\alpha_{e}^{d}(\omega)\right]{\cal E}^2,
\end{eqnarray} 
where the subscripts \textit{g} and \textit{e} stand for the ground and excited states. In our work, we identify the magic wavelengths by 
plotting the dynamic dipole polarizabilities against $\omega$ values and finding out their crossings for the ground and the excited states. The 
two polarizability curves usually cross in between of any two resonant transitions of the the states and the $\omega$ values at those crossings 
are referred to as the magic wavelengths.

\subsubsection{Quadrupole polarizability}

The quadrupole Stark shift in the energy level of an atomic system in state $|\Psi_n\rangle \equiv |\gamma _n J_n M_{J_n}\rangle$ due to the 
electric field gradient at frequency $\omega$ is given by second term of Eq. (\ref{eqshift}). In the sum-over-states approach,
these polarizabilities are given by
\begin{eqnarray}
\alpha^q_{n}(\omega)&=&=-\sum_{m\neq n}{(q^*)_{nm}(q)_{mn}} \times \left[\frac{1}{\delta E_{nm}+\omega}+\frac{1}{\delta E_{nm}-\omega}\right].\label{eq-quad}
\end{eqnarray}
Here $(q)_{mn}=\langle \Psi _m|Q|\Psi _n\rangle$ is the E2 matrix element  and $\delta E_{nm} = E_n^0-E_m^0$ being the unperturbed energies of the 
corresponding states as mentioned above. For linearly polarized light, we can conveniently evaluate $\alpha_{n}^q(\omega)$ by rewriting it in terms 
of different tensor components as~\cite{sukhjit2018}
\begin{eqnarray}
\alpha^q_{n}(\omega)&=&\alpha_{n}^{q(0)}(\omega) +  \frac{3M_{J_n}^2-J_n(J_n+1)}{J_n(2J_n-1)} \alpha_{n}^{q(2)}(\omega)+\frac{3(5M_{J_n}^2-J_n^2 -2J_n)(5M_{J_n}^2+1-J_n^2)-10M_{J_n}^2(4M_{J_n}^2-1)}{J_n(J_n-1)(2J_n-1)(2J_n-3)} \alpha_{n}^{q(4)}(\omega), \nonumber\label{alphat} \\
\end{eqnarray}
where $\alpha_{n}^{q(0)}(\omega)$, $\alpha_{n}^{q(2)}(\omega)$ and $\alpha_{n}^{q(4)}(\omega)$ are defined as the scalar, tensor component of rank 
2 and tensor component of rank 4 of $\alpha^q_{n}(\omega)$, respectively, of the quadrupole polarizability. The corresponding expressions for different components can be given 
in terms of $\langle \gamma _nJ_n||{\bf Q}||\gamma _m J_m \rangle$ reduced E2 matrix elements by
\begin{eqnarray}
\alpha_{n}^{q(0)}(\omega)&=& \sum_{m \ne n} W_{n}^{q(0)} |\langle \gamma _nJ_n||{\bf Q}||\gamma _m J_m \rangle|^2 \left [\frac{ 1}{\delta E_{nm} +\omega}+\frac{ 1}{\delta E_{nm}-\omega}\right],\label{eqpolz1q} \\
 \alpha_{n}^{q(2)}(\omega)&=& \sum_{m \ne n} W_{n,m}^{q(2)} |\langle \gamma _nJ_n||{\bf Q}||\gamma _m J_m \rangle|^2  \left [\frac{ 1}{\delta E_{nm} +\omega}+\frac{1}{\delta E_{nm}-\omega}\right] \label{eqpolz2q}
\end{eqnarray}
and
\begin{eqnarray}
\alpha_{n}^{q(4)}(\omega)&=& \sum_{m \ne n} W_{n,m}^{q(4)} |\langle \gamma _nJ_n||{\bf Q}||\gamma_m J_m \rangle|^2  \left [\frac{1}{\delta E_{nm}+\omega}+\frac{ 1}{\delta E_{nm}-\omega} \right ]. \label{eqpolz3q}
\end{eqnarray}
In the above expressions, the coefficients are given by
\begin{eqnarray}
W_{n}^{q(0)} &=&-\frac{1}{5(2J_n+1)}, \label{w1} \\
W_{n,m}^{q(2)}&=&-\sqrt{\frac{10J_n(2J_n-1)}{7(J_n+1)(2J_n+1)(2J_n+3)}} (-1)^{J_n+J_m+1}  \left\{ \begin{array}{ccc}
                             J_n& 2 & J_n\\
                          2 & J_m &2 
                         \end{array}\right\} \label{w2}
\end{eqnarray}
and
\begin{eqnarray}
W_{n,m}^{q(4)} &=&\sqrt{\frac{J_n(J_n-1)(2J_n-1)(2J_n-3)}{70(2J_n+1)(J_n+1)(J_n+2)}}  \frac{9(-1)^{J_n+J_m+1}}{\sqrt{(2J_n+5)(2J_n+3)}}
                \left\{ \begin{array}{ccc}
                                            J_n& 4 & J_n\\
                                            2 & J_m & 2 
                                           \end{array}\right\}. \label{w3} \ \ \ \ \ \                                       
\end{eqnarray}
The above expressions can be used to evaluate both the static ($\omega = 0$) and dynamic ($\omega \neq 0$) quadrupole polarizabilities. Only the 
scalar component will contribute to the states with angular momentum $J=1/2$, while both the scalar and tensor components with rank 2 will contribute 
to the states with $J=3/2$ and the tensor quadrupole polarizability corresponding to the rank 4 appears for states with $J=5/2$. It can be followed 
from this is that it is more laborious and challenging to obtain accurate values of the quadrupole polarizabilities of the $D$ states as they will 
need many more matrix elements involving a large number of intermediate states. According to the Eq.~(\ref{alphat}), the total polarizabilities for 
the states with $J=3/2$ is evaluated as $\alpha^{q}_{n}(\omega)=\alpha^{q(0)}_{n}(\omega)-\alpha^{q(2)}_{n}(\omega)$ for $M_J=1/2$ and as 
$\alpha^{q}_{n}(\omega)=\alpha^{q(0)}_{n}(\omega)+\alpha^{q(2)}_{n}(\omega)$ for $M_J= 3/2$. The total polarizability for the states with $J=5/2$ can be 
calculated as $\alpha^{q}_{n}(\omega)=\alpha^{q(0)}_{n}(\omega)-0.8 \alpha^{q(2)}_{n}(\omega)+4 \alpha^{q(4)}_{n}(\omega)$ for $M_J=\pm 1/2$, 
as $\alpha^{q}_{n}(\omega)=\alpha^{q(0)}_{n}(\omega)-0.2 \alpha^{q(2)}_{n}(\omega)-6 \alpha^{q(4)}_{n}(\omega)$ for $M_J=\pm 3/2$ and 
as $\alpha^{q}_{n}(\omega)=\alpha^{q(0)}_{n}(\omega)+\alpha^{q(2)}_{n}(\omega)+2\alpha^{q(4)}_{n}(\omega)$ for $M_J=\pm 5/2$.

\subsection{Procedure adopted for polarizability evaluation}

Each component of dipole as well as quadrupole polarizabilities in the considered alkaline earth metal ions can be conveniently estimated by 
calculating contributions separately due to the core, core-valence and valence correlations~\cite{arora2012,kaur2015}. These can be written as
\begin{eqnarray}
\alpha_{n}^{o}(\omega)  &=& \alpha_{n,c}^{o}(\omega) + \alpha_{n,cv}^{o}(\omega) + \alpha_{n,v}^{o}(\omega) ,
\label{eq26}
\end{eqnarray}
where $o$ denotes either $d$ or $q$ for the E1 or E2 polarizability respectively. The $\alpha_{n,c}^{o}(\omega)$, $\alpha_{n,cv}^{o}(\omega)$ and 
$\alpha_{n,v}^{o}(\omega)$ present the core, core-valence and valence correlation effects respectively. The most important contributions to different 
components of polarizability arises through $\alpha_{n,v}^{o}(\omega)$, since it takes into account the dominant correlation effects from the 
calculations of atomic wave functions. For calculating the $\alpha_{n,v}^{o}(\omega)$ contribution, we first calculate the wave functions of atomic 
states and matrix elements of as many as low-lying states of the considered alkaline earth metal ions using relativistic all-order method as already 
explained earlier in this section. The dominant contribution to $\alpha_{n,v}^{o}(\omega)$ is evaluated by combining these matrix elements with the 
experimental energies listed in the the National Institute of Science and Technology database (NIST)~\cite{kramida}. This is called as the ``Main'' 
contribution to $\alpha_{n,v}^{o}(\omega)$, whereas the smaller contributions from the high-lying excited states are evaluated using the DHF method 
and are termed as the ``Tail'' contribution to $\alpha_{n,v}^{o}(\omega)$. The smaller contributions arising from the valence-core correlation 
contribution $\alpha_{n,cv}^{o}(\omega)$ is evaluated using DHF method and from the core correlation contribution $\alpha_{n,c}^{o}(\omega)$ have been evaluated using 
the random phase approximation (RPA).

\section{Data analysis and discussion} \label{results}

\subsection{Reduced E2 matrix elements} 

The reduced E2 matrix elements calculated using the DHF, SD, SDsc, SDpT and SDpTsc methods for a number of $S-D$, $D-G$, $P-P$ and $P-F$ transitions 
of the Mg$^+$, Ca$^+$, Sr$^+$ and Ba$^+$ alkaline-earth metal ions are listed in Tables \ref{matrixmge2}, \ref{matrixcae2}, \ref{matrixsre2} and
\ref{matrixbae2}, respectively. We also list our recommended values along with the uncertainties for the considered transitions in the same tables. 
For the range $0.5<R<1.5$, the uncertainties are determined as the maximum difference between the final value of matrix element and the other three 
all-order values. However for $1.5<R<3$, the uncertainties are calculated as max(SDsc - SD, SDsc - SDpT, SDsc - SDpTsc). Also, if $R>3$, the 
uncertainties are given as max(SDsc - SDpT, SDsc - SDpTsc).  We have calculated these E2 matrix elements for about 114 transitions in Mg$^+$, 
114 transitions in Ca$^+$, 130 transitions in Sr$^+$ and 96 transitions in Ba$^+$.

\subsection{E2 transition probabilities}

By using our recommended values of the E2 matrix elements and experimental values for the wavelengths, we have calculated the line strengths $S_{vk}$, transition probabilities 
$A_{vk}$ and oscillator strengths $f_{kv}$ for all the considered E2 transitions of Mg$^+$, Ca$^+$, Sr$^+$ and Ba$^+$ which are presented in Tables 
\ref{parametersmge2}, \ref{parameterscae2}, \ref{parameterssre2} and \ref{parametersbae2} respectively. The experimental values for 
the wavelengths were derived from the excitation energies listed in the NIST database~\cite{kramida}. This includes $nS_{1/2}-n'D_{5/2,3/2}$, $n'D-n''D$,
$n'D_{5/2,3/2}-mG_{9/2,7/2}$, $qP-q'P$ and $qP_{3/2,1/2}-m'_{7/2,5/2}F$ E2 transitions with $\{n,n',n''\}= 3-6$, $m=5,6$, $\{q,q'\}= 3-5$ and $m'=4-5$ in Mg$^+$; 
$n=4-7$, $\{n',n''\}= 3-6$, $m=5,6$, $\{q,q'\}= 4-6$ and $m'=4-5$ in Ca$^+$; $n=5-8$, $\{n',n''\}= 4-7$, $m=5,6$, $\{q,q'\}= 5-8$ and $m'=4-5$ in 
Sr$^+$; and $n=6-9$, $\{n',n''\}= 5-8$, $m=5,6$ and $\{q,q'\}= 6-8$ in Ba$^+$. We have made comparison of the our evaluated data with 
the previously available literature below. 

In Table \ref{comparmge2}, a comparison of our calculated data for Mg$^+$ is presented with previously available literature. Comparison with the NIST database indicates very good agreement for all the available transitions except for the $3D_{5/2}-3S_{1/2}$ and $4D_{5/2}-3S_{1/2}$ transitions. 
However, our evaluated results for these transitions are in excellent agreement with the WBEMPT calculations by Celik et al.~\cite{celik}, RCC 
calculations by Majumder et al.~\cite{sanjoy} and MCHF calculations by Fischer et al.~\cite{fischer}, which expresses the reliability in our 
calculations for these transitions despite of differences with NIST database. Celik et al.~\cite{celik}, Majumder et al.~\cite{sanjoy} and Fischer
et al.~\cite{fischer} have also provided calculations of transition probabilities of many other transitions. It is reflected in the table that, in 
general our evaluated data is in very good agreement with the available literature except for a few discrepancies discussed next. For the 
$3P_{3/2}-3P_{1/2}$ transition a good agreement of our value is seen with ~\cite{celik} and ~\cite{sanjoy}, whereas a small discrepancy can be noticed 
with the MCHF results in Ref.~\cite{fischer}. For the $4P-3P$ transitions our results match extremely well with the results in Ref.~\cite{fischer} as 
compared to slight difference with those presented in Refs.~\cite{celik,sanjoy}, whereas, for the $5P_{3/2}-5P_{1/2}$ transition our value is much 
closer to the values in Ref.~\cite{celik}. It can be noticed that our relativistic all-order calculations for the $5P-4F$ and $3D_{3/2}-3D_{5/2}$ 
transitions present major discrepancy with the RCC calculations of Majumder et al.~\cite{sanjoy}. The reason could be we have used here B-spline 
basis functions, whereas Majumder et al.~\cite{sanjoy} have used Gaussian type orbitals (GTOs) to describe the orbitals. Unless, suitable optimized parameters 
are used to describe $4F$ orbitals GTOs do not generate enough bound orbitals. Since we have imposed boundary conditions with a large cavity 
radius and our bound orbitals obtained using the numerical approach, we assume our results for the $4F$ orbitals are more reliable. The comparison of 
the transition probability value for the $5D_{3/2}-3D_{5/2}$ transition depicts that our value matches neither with the values presented in 
Ref.~\cite{celik} nor with Ref.~\cite{sanjoy}. For this transition, the values of Ref.~\cite{celik} and Ref.~\cite{sanjoy} also show mutual conflict 
with each other. Such disagreements could be partly because of the aforementioned reason. In addition, electron correlation effects arising through 
non-linear terms and triple excitations that are neglected in our all-order theory contribute significantly to the states with higher angular 
momentum \cite{bijaya2016}. Inclusion of these contributions in the calculations may remove most of these discrepancies.

In Table~\ref{parameterse2}, we have presented the comparison of the transition probabilities $A_{vk}$ and oscillator strengths $f_{kv}$ for 
the $nD-(n+1)S$ and $nD_{5/2}-nD_{3/2}$ transitions of Ca$^+$($n=3$), Sr$^+$($n=4$) and Ba$^+$($n=5$) with available literature. In our previous paper
\cite{mandeep}, we had presented a few E2 matrix elements by employing the SD method approximation. Here, in present paper we have used the recommended values of the 
matrix elements based on the criteria mentioned in Sec~\ref{matrix}. In this table, we have also presented the comparison of our data with the values 
available in the NIST database. In literature, values for the E2 transition probabilities are available only for the $3D-4S$ transitions in Ca$^+$ \cite{guan2015,filippin,shao2017}; $4D-5S$ transitions 
in Sr$^+$~\cite{filippin}; the $5D-6S$ transitions in Ba$^+$~\cite{filippin,gurell2007} and the $3D_{5/2}-3D_{3/2}$~\cite{filippin}, $4D_{5/2}-4D_{3/2}$~\cite{filippin} and $5D_{5/2}-5D_{3/2}$~\cite{filippin,gurell2007} 
transitions in Ca$^+$, Sr$^+$ and Ba$^+$ respectively. We notice that discrepancy is found for the results in Ca$^+$ whereas, our results for Sr$^+$ exhibit excellent 
agreement with values from NIST database~\cite{kramida}. More importantly our calculated quadrupole transition probability for the $3D_{5/2}-4S_{1/2}$ 
transition in Ca$^+$ is in excellent agreement with the recent experimental result reported by Shao et al.~\cite{shao2017}. This strongly advocates 
for the accuracy of our results. Also, our estimated values for Ca$^+$ exhibit good agreement with the recent semi-empirical core potential 
calculations by Filippin et al.~\cite{filippin} and RCC calculations by Guan et al.~\cite{guan2015}. Similarly, the transition probabilities of 
Sr$^+$ and Ba$^+$ are in good agreement with the results provided by Filippin et al.~\cite{filippin}. For Ba$^+$, our values show good agreement with 
the pseudo-relativistic Hartree-Fock method calculations by Gurell et al.~\cite{gurell2007}. 

\subsection{Static Dipole polarizability}

Using our previously reported E1 matrix elements in Ref. \cite{mandeep} and the experimental energies taken from the NIST database, we evaluate the static dipole 
polarizabilities of the ground and excited states of Mg$^+$, Ca$^+$, Sr$^+$ and Ba$^+$ ions. Then, we compare these values with the previously available
experimental and theoretical results in Tables \ref{staticmg}, \ref{staticca}, \ref{staticsr} and \ref{staticba}. The scalar polarizabilities of the 
considered ground and excited states along with the tensor polarizabilities for the states with $J > 1/2$ are presented separately. The
contributions from the `Main' and `Tail' parts of the valence correlations along with the contributions from the core-valence and core correlations 
are also given in the same tables. We discuss these results for the individual ion below.

\subsubsection{Mg$^+$}

Table \ref{staticmg} presents the tabulation of the static polarizability values for the $(3-5)S_{1/2}$, $(3-5)P_{3/2,1/2}$ and 
$(3-4)D_{5/2,3/2}$ states of Mg$^+$. Along with this, the comparison of  our results for the $3S_{1/2}$, $3P_{3/2,1/2}$ and $3D_{5/2,3/2}$ states 
with the values provided by Mitroy et al.~\cite{mit2009} is made in the same table. The static polarizability values by Mitroy et al.~\cite{mit2009} 
are computed using a semi-empirical single electron analysis combined with the MBPT-SD method. It can clearly be noticed that our ground state 
polarizability values are in excellent agreement with their values. A similar trend of remarkable agreement can be seen for the scalar as well as 
tensor polarizability values of the $3P$ states as well as of the $3D$ states between our and their values. Hence, the accuracy of our results for 
excited states can also be ascertained from this. The literature values for the static polarizabilities for other excited states are not available 
and these calculations have been carried our for the first time in this work.

\subsubsection{Ca$^+$}

As can be seen that Table \ref{staticca} enlists the values of the static polarizabilities for the states $(4-6)S_{1/2}$, $(4-6)P_{3/2,1/2}$ and 
$(3-5)D_{5/2,3/2}$ states of Ca$^+$. Our estimated values of the polarizabilities for the $4S$, $4P$ and $3D$ states are in excellent agreement with 
the MBPT-SD calculations by Safronova et al.~\cite{safronova2011}. Our calculated value of 76.0$\pm$0.6 a.u. for the $4S_{1/2}$ state is in close agreement 
with the other theoretical calculations that is using the relativistic structure model by Tang et al.~\cite{tang13} and diagonalizing semi-empirical 
Hamiltonian calculations by Mitroy et al.~\cite{mitroy2008}, which are 75.28 a.u. and 75.49 a.u. respectively. Along with the ground state results, 
our calculated polarizability values for the $3D_{3/2}$ and $3D_{5/2}$ states are also compared with Mitroy et al.~\cite{mitroy2008} and Tang et 
al.~\cite{tang13}, which represent very good agreements. The comparison of our calculated polarizability values for the $4S_{1/2}$, $3D_{3/2}$ and  
$3D_{5/2}$ states show reasonable agreement within error bars with {\it ab initio} RCC calculations by Sahoo et al.~\cite{sahoo2009a}. Also, our estimated  
$4S_{1/2}$ state polarizability value is very close to the experimental result, obtained with spectral analysis technique  by Chang 
et al.~\cite{chang83}, within the uncertainty limits. However, large discrepancy can be noticed between our values of the scalar polarizabilities for 
the $4P$ states with Tang et al. and Mitroy et al.~\cite{tang13,mitroy2008}. To our knowledge, no data is available in the literature for comparison 
of the the static polarizability of other excited states that are calculated in the present work.

\subsubsection{Sr$^+$}

In this subsection, we compare our static polarizability results for the Sr$^+$ ion tabulated in Table \ref{staticsr} for the $(5-8)S_{1/2}$, 
$(5-7)P_{3/2,1/2}$ and $(4-6)D_{5/2,3/2}$ states. Our estimated value for the $5S_{1/2}$ state is in very good agreement with the relativistic 
RCC calculations given by Sahoo et al.~\cite{sahoo2009b}, relativistic all-order values by Jiang et al. and Safronova~\cite{dansha2009,safr2010} and 
semi-empirical calculations by Mitroy et al.~\cite{mitroy2008PRA}. Along with the theoretical results, the ground state polarizability is also in 
agreement within error bars with the measurements by Barklem et al. and Nunkaew et al.~\cite{barklem,nunkaew2009}. However, a considerable discrepancy
of our polarizabilities for the $4D$ states can be noticed with the values reported by Barklem et al.~\cite{barklem}. The origin of this discrepancy 
is mainly from the fact that core contribution is omitted in their calculations. Relativistic all-order calculations by Safronova et al.~\cite{safr2010}
are available for the $(6-8)S$, $(5-7)P$ and $(4-6)D$ states and our values for these states exhibit good accord with them. Mitroy et 
al.~\cite{mitroy2008PRA} have also given polarizabilities for the $6S$, $(5-6)P$ and $4D$ states which show slight disagreement with our values. 
However, our values are produced by using the relativistic all-order method and expected to be more accurate. Our polarizability values for the 
$4D$ states are in agreement within uncertainty limits with the results reported in Refs.~\cite{sahoo2009b,dansha2009,jiang2016}. 

\subsubsection{Ba$^+$}

Table \ref{staticba} tabulates the static polarizability results for the $(6-8)S_{1/2}$, $(6-8)P_{3/2,1/2}$ and $5D_{5/2,3/2}$ states of Ba$^+$. For 
the ground state, our polarizability value of 124$\pm$2 a.u. is an excellent agreement with the high-precision measurement presented by Snow et 
al~\cite{snow2007} based on the resonant excitation Stark ionization spectroscopy and with the radio frequency resonance measurement by Gallagher 
et al~\cite{gallagher1982}. Also, our ground state polarizability value for ground state is in very good agreement with the value reported by 
Safronova et al.~\cite{isk2008}. Along with this, our values for the $6S_{1/2}$, $5D_{3/2}$ and $5D_{5/2}$ states are also compared with the RCC 
calculations by Sahoo et al.~\cite{sahoo2009b} and Barrett et al~\cite{barrett2019}. For ground state, the agreement is reasonable between our and 
their reported values. Similarly, for the $5D_{5/2}$ state, our values match well with Sahoo et al. and Barrett et al.~\cite{sahoo2009b,barrett2019} 
within uncertainty limits. However, our polarizability value for the $5D_{3/2}$ state lies just outside the uncertainty limits of the value by Sahoo 
et al.~\cite{sahoo2009b}. Also, the non availability of the literature for the static polarizability values of other excited states restricts us from
making comparison of our values for those states.

\subsection{Static quadrupole polarizability}

In this section we present the static quadrupole polarizabilities of Mg$^+$, Ca$^+$, Sr$^+$ and Ba$^+$ ions in Tables \ref{staticquadmg}, 
\ref{staticquadca}, \ref{staticquadsr} and \ref{staticquadba}, respectively. We give the scalar polarizabilities of the ground and excited states 
as well as the tensor rank 2 component of polarizabilities for the states with $J > 1/2$ of the above ions. In this case, there are also contributions
to the states with $J = 5/2$ \cite{sukhjit2018}. Also, contributions from `Main' and `Tail' of the valence correlations, and core-valence and core 
correlations are enlisted in the same tables for all the considered ions. We have used our recommended set of E2 matrix elements and the experimental 
energies available in the NIST database to determine the `Main' contributions. As can be seen from the above tables, the valence-core and `Tail' 
contributions towards the total polarizabilities are very small in all investigated states of the ions. This is why we have evaluated them by using 
the DHF method. Since the core correlation contributions are little large, we have estimated them using RPA to account for the core-polarization 
correlation effects to all-order. We discuss the quadrupole polarizability values below for each ion.

\subsubsection{Mg$^+$}

Table \ref{staticquadmg} enlists the quadrupole static polarizability values for the $3S_{1/2}$, $3P_{3/2,1/2}$ and $3D_{5/2,3/2}$ states of Mg$^+$. 
Its core correlation contribution is estimated to be 0.52$\pm$0.03 a.u. using RPA. As can be seen from the table that there are not much studies of these 
quantities are available in the literature for Mg$^+$ to make rigorous comparison of our calculations. The only values available in the literature 
are for the polarizability of the $3S_{1/2}$ state. Our value 156.0$\pm$0.4 a.u. for this ground state agrees perfectly with the values 156.1 a.u. 
reported by Mitroy et al.~\cite{mitroy2008} using the  diagonalization of semi-empirical Hamiltonian in a large-dimension single-electron basis. 
Also, an excellent agreement can be noticed between our value and the value 156.02$\pm$1.27 a.u. reported by Sahoo et al.~\cite{sahoo2007} using the RCC method. 

\subsubsection{Ca$^+$}

We present the static quadrupole polarizabilities of the $4S_{1/2}$, $4P_{3/2,1/2}$ and $3D_{5/2,3/2}$ states of Ca$^+$ in Table \ref{staticquadca}. The 
RPA value for the core correlation came out to be 6.9$\pm$0.3 a.u.. The E2 polarizability values for the  $4S_{1/2}$ and $4P_{1/2}$ states come only from 
the scalar component, whereas both the scalar as well as tensor rank 2 components are present in the E2 polarizabilities of the $4P_{3/2}$ and 
$3D_{5/2,3/2}$ states. However, the E2 polarizability of the $3D_{5/2}$ state can also have contribution from the tensor rank 4 component.
It can clearly be noticed that our value for the ground state quadrupole polarizability, 874$\pm$9 a.u., is in excellent agreement with the relativistic 
all-order value 871$\pm$4 a.u. reported by Safronova et al.~\cite{safronova2011} as well as with value 875.1 a.u. reported by Mitroy and 
Zhang~\cite{mitroy2008}. No other data is available in the literature to compare our values for the $4P_{3/2,1/2}$ and $3D_{5/2,3/2}$ states. The 
values of the quadrupole polarizabilities reported for the $4S$ and $3D$ states in our previous work~\cite{sukhjit2018} were carried out using 
the SD and SDpT methods, whereas in present work we have used the recommended matrix elements based on criteria mentioned in Sec. \ref{matrix} 
to improve their accuracies.

\subsubsection{Sr$^+$}

Table \ref{staticquadsr} summarizes the static quadrupole polarizabilities of the $5S_{1/2}$, $5P_{3/2,1/2}$ and $4D_{5/2,3/2}$ states of Sr$^+$.
The core correlation contribution is estimated using RPA as 17.1$\pm$0.9 a.u.. Our quadrupole polarizability value 1375$\pm$16 a.u. of the ground state differs 
by 0.9 percent from the value of 1346 a.u. reported by Mitroy and Zhang~\cite{mitroy2008PRA}. Whereas, the comparison of our calculation with the value 
calculated by Jiang et al. ~\cite{jiang2016} obtained using the relativistic semi-empirical approach and Safronova~\cite{safr2010} calculated using 
the relativistic all-order method show very good agreements. Jiang et al..~\cite{jiang2016} have also reported the scalar components of the 
quadrupole polarizability for the $5P_{3/2,1/2}$ and $4D_{5/2,3/2}$ states. A very good agreement between our results for $5P_{3/2,1/2}$ with ~\cite{jiang2016} is noticed.
However, the difference of approximately 1.8 percent can be found 
between our and their values for the $4D_{3/2}$ state whereas the polarizability value for $4D_{5/2}$ state matches within error bars with them. Our ground state polarizability value shows significant difference with the experimental value reported by 
Nunkaew et al.~\cite{nunkaew2009}, who have used indirect spin-orbit K splittings techniques to obtain their result. Such discrepancy was also seen from another theoretical work by Jiang et al~\cite{jiang2016}. 

\subsubsection{Ba$^+$}

The static scalar quadrupole polarizability values for the $6S_{1/2}$ and $5D_{5/2,3/2}$states of Ba$^+$ are tabulated in Table \ref{staticquadba}. 
The tensor contributions to the E2 polarizabilities of the $5D_{5/2,3/2}$ states are also illustrated in the same table. The RPA value for the core 
contributions is obtained as 46$\pm$2 a.u. for Ba$^+$. Many calculations of the ground state quadrupole polarizability of this ion are reported by 
different groups by using a variants of many-body methods. Our ground-state scalar quadrupole polarizability value of came out to be 4192$\pm$47 a.u., which is in perfect agreement with the value  of 4182$\pm$34 a.u. within uncertainty limits reported by Iskrenova et al.~\cite{isk2008}. However, a slight difference can be noticed between our value and the value reported by Safronova~\cite{uisaf2010}. The corresponding experimental result is reported as 4420$\pm$250 a.u. by Snow et al.~\cite{snow2007} and it is very clear that our value lies within the uncertainty of this measurement. The value of our scalar quadrupole polarizability for the $5D_{3/2}$ state lies very well within the uncertainty limits of the value provided by Sahoo et al. ~\cite{sahoo2012}, whereas the value for the $5D_{5/2}$ and $6S_{1/2}$ states are found to be close to the uncertainty limits. Hence, the accuracy of our results is emphasized as to be reliable. 

\subsection{Magic and tune-out wavelengths}\label{magic}

We now discuss here magic and tune-out wavelengths inferred for the Mg$^+$, Ca$^+$, Sr$^+$ and Ba$^+$ ions. For this purpose, we have evaluated 
the dynamic E1 polarizabilities of the atomic states of interest for a wide range of wavelengths. Since these dynamic polarizabilities are 
evaluated by adopting the same procedure as the static E1 polarizabilities that were discussed in the previous sub-section, we anticipate that 
the dynamic E1 polarizabilities are of similar accuracies with the static values. As mentioned earlier, we have assumed that the external 
electric field is linearly polarized for the determination of these E1 polarizabilities, and at the $\lambda_{\rm{T}}$ values the dynamic E1 
polarizabilities of a given atomic state vanish while the differential dynamic E1 polarizabilities of an atomic transition nullify at the 
$\lambda_{\rm{magic}}$ values. We present both the $\lambda_{\rm{T}}$ and $\lambda_{\rm{magic}}$ values for each of the considered ion below. 

\begin{figure}[t]
    \centering
\begin{tabular}{cc}.
      \includegraphics[height=8cm,width=8.5cm]{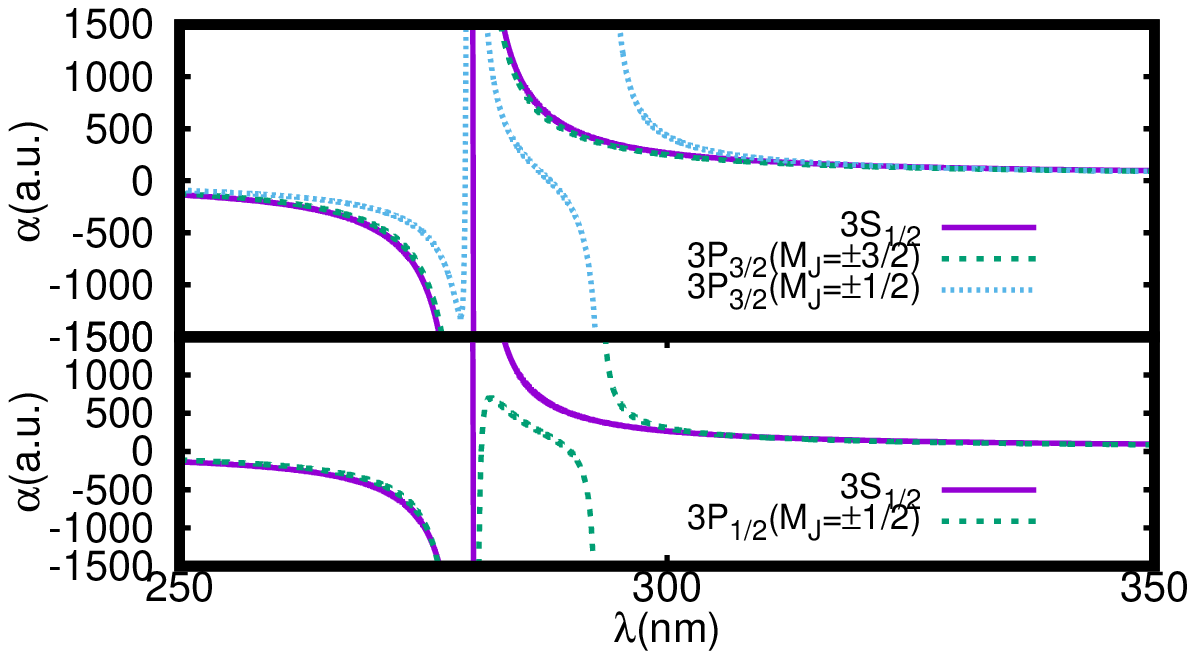} & \includegraphics[height=8cm,width=8.5cm]{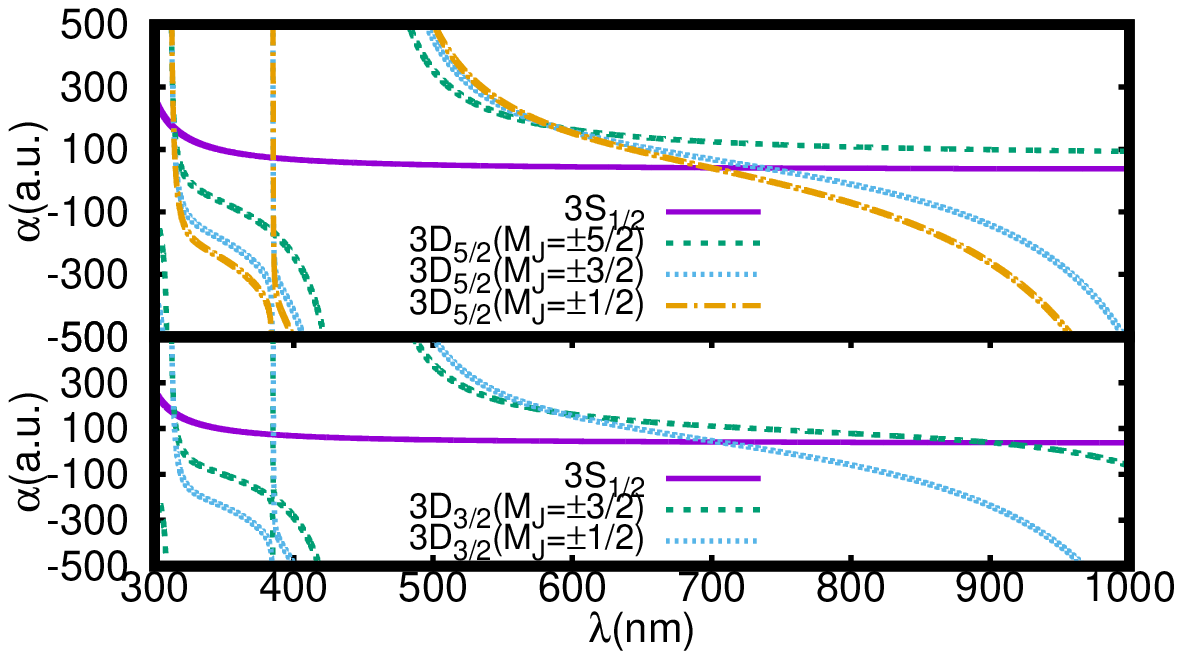} \\
    (a) & (b) \\
\end{tabular}
    \caption{Plots depicting dynamic E1 polarizabilities of (a) the ground $3S_{1/2}$ and excited $3P_{3/2,1/2}$ states, and (b) the ground $3S_{1/2}$ 
    and excited $3D_{5/2,3/2}$ states in Mg$^+$ for the linearly polarized light. The crossings of polarizability curves  between two resonances correspond to the 
    magic wavelengths.}  \label{mgmagic}   
\end{figure}

\subsubsection{Mg$^+$}

In Table \ref{magicmg}, we list the magic wavelengths for the $3S_{1/2}-3P_{3/2,1/2}$ and $3S_{1/2}-3D_{5/2,3/2}$ transitions of Mg$^+$ at different 
$M_J$ sublevels. These values are inferred from the plots of dynamic polarizabilities as shown in Figs.~\ref{mgmagic}(a) and \ref{mgmagic}(b). As can be seen 
from the table and figures that the magic wavelength at 276.1$\pm$0.7 nm has large negative polarizability which is $-1266$ a.u.. It means that it can support blue detuned trap 
to conduct an experiment at this wavelength to reduce systematics due to the Stark effect. Similarly, the magic wavelengths at the 307$\pm$2 nm 
for the $3S_{1/2}-3P_{1/2}$ transition and at the 324$\pm$3 nm for the $3S_{1/2}-3P_{3/2}$ transition support the red detuned trap. Also 
six $\lambda_{\rm{magic}}$ values are observed for the $3S_{1/2}-3D_{3/2}$ transition between different resonances. Out of them two at 385.011$\pm$0.003 nm and 
385.376$\pm$0.003 nm are found to be very close to the $3D_{3/2}-5P_{1/2}$ resonance. For the magic wavelengths at 705$\pm$1 nm and 901$\pm$3 nm, the polarizabilities
are very small which make these wavelengths of less practical importance.  Again, for the  $3S_{1/2}-3D_{5/2}$ transition, seven magic wavelengths are
found. Out of these, three magic wavelengths lie very close to the $3D_{5/2}-5F_{5/2}$ transition and two other at 385.121$\pm$0.003 nm and 385.170$\pm$0.003 nm are in the close vicinity of resonant transition $3D_{5/2}-5P_{3/2}$. It can be noticed that most of the magic wavelengths of Mg$^+$ support very shallow red detuned traps. 

\begin{figure}[t]
    \centering
\begin{tabular}{cc}.
      \includegraphics[height=8cm,width=8.5cm]{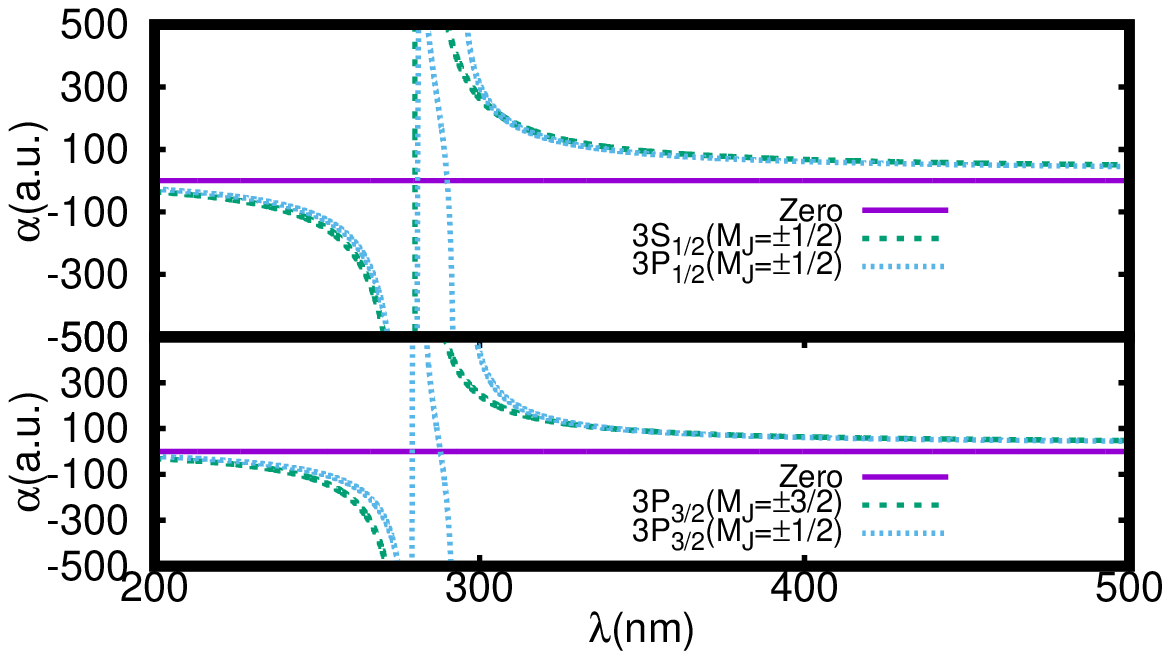} & \includegraphics[height=8cm,width=8.5cm]{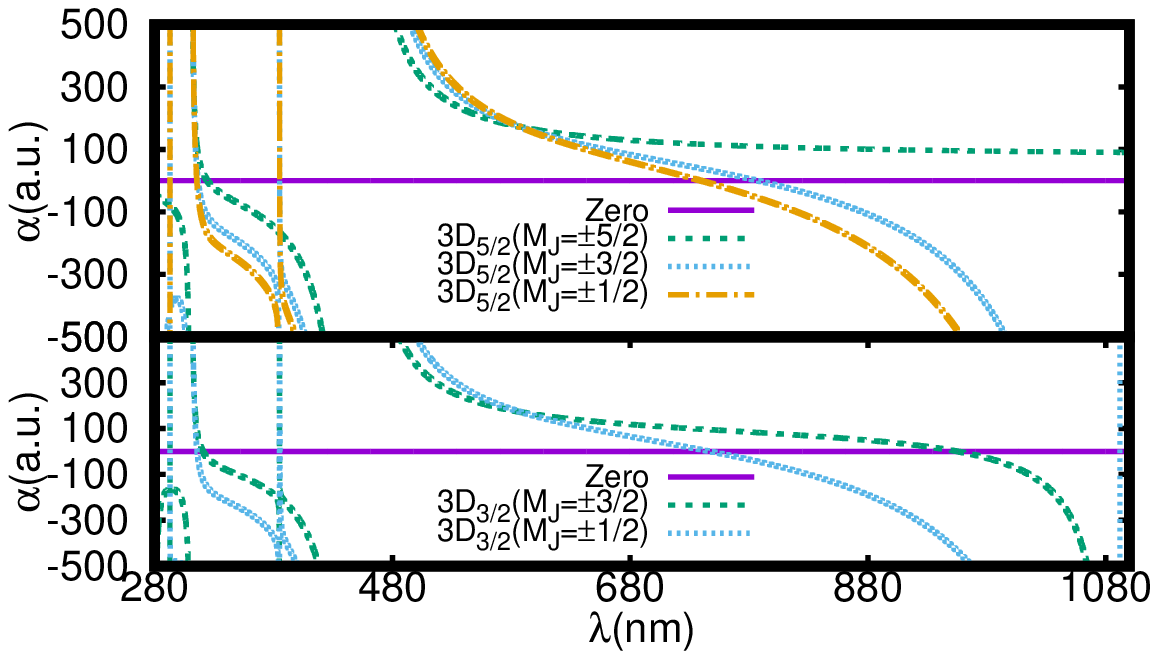}\\
    (a) & (b) \\
\end{tabular}
    \caption{Plots for the dynamic E1 polarizabilities of (a) the ground $3S_{1/2}$ and $3P_{3/2,1/2}$ states, and (b) the $3D_{5/2,3/2}$ states 
    in Mg$^+$ for the linearly polarized light. The crossings of the dynamic dipole polarizability curves with zero  depict
    the tune-out wavelengths.}  \label{mgtune}  
\end{figure}

In Table \ref{tuneout}, we have illustrated the tune-out wavelengths for the $nS_{1/2}$, $nP_{3/2,1/2}$ and $nD_{5/2,3/2}$ states of Mg$^+$ with $n=3$.
These values are inferred from Figs. \ref{mgtune}(a) and \ref{mgtune}(b). We have presented the cancellation of all contributions to the polarizabilities 
of the $3S_{1/2}$ and $3P_{3/2,1/2}$ states at $\lambda_{\rm{T}}$ in Fig. \ref{mgtune}(a). The $\lambda_{\rm{T}}$ values at 280.110$\pm0.009$ nm and 
at 280.95$\pm0.04$ nm of the $3S_{1/2}$ and $3P_{1/2}$ states, respectively, lie close to the  $3P_{1/2}-3S_{1/2}$ resonance. For the $3P_{1/2}$ state,
the $\lambda_{\rm{T}}$ value at 290.05$\pm$0.06 nm can be found between the $3P_{1/2}-3S_{1/2}$ and $3P_{1/2}-4S_{1/2}$ resonances. All these tune-out
wavelengths including those for the $3P_{3/2}$ state lie in ultraviolet region so are of less practical use. Further, Fig. \ref{mgtune}(b) exhibits 
the tune-out wavelengths for the $3D_{5/2,3/2}$ states. For the $3D_{3/2}$ state, the $\lambda_{\rm{T}}$ values at 385.04$\pm$0.01 nm and 
385.42$\pm0.01$ nm are very near to the $3D_{3/2}-5P_{1/2}$ resonance whereas the one at 747$\pm2$ nm lie well within the visible region. Hence, the 
latter one appears to gain the experimental importance. The $\lambda_{\rm{T}}$ at 955$\pm2$ nm falls in the infrared region. Similarly for the 
$3D_{5/2}$ state, the tune-out wavelengths at 385.17$\pm0.01$ nm and 385.21$\pm0.01$ nm are in the close vicinity of the $3D_{5/2}-5P_{3/2}$ 
transition at 384.93 nm, whereas $\lambda_{\rm{T}}$s at 739$\pm2$ nm lies in the visible  region.  

\subsubsection{Ca$^+$}

\begin{figure}[t]
    \centering
\begin{tabular}{cc}.
      \includegraphics[height=8cm,width=8.5cm]{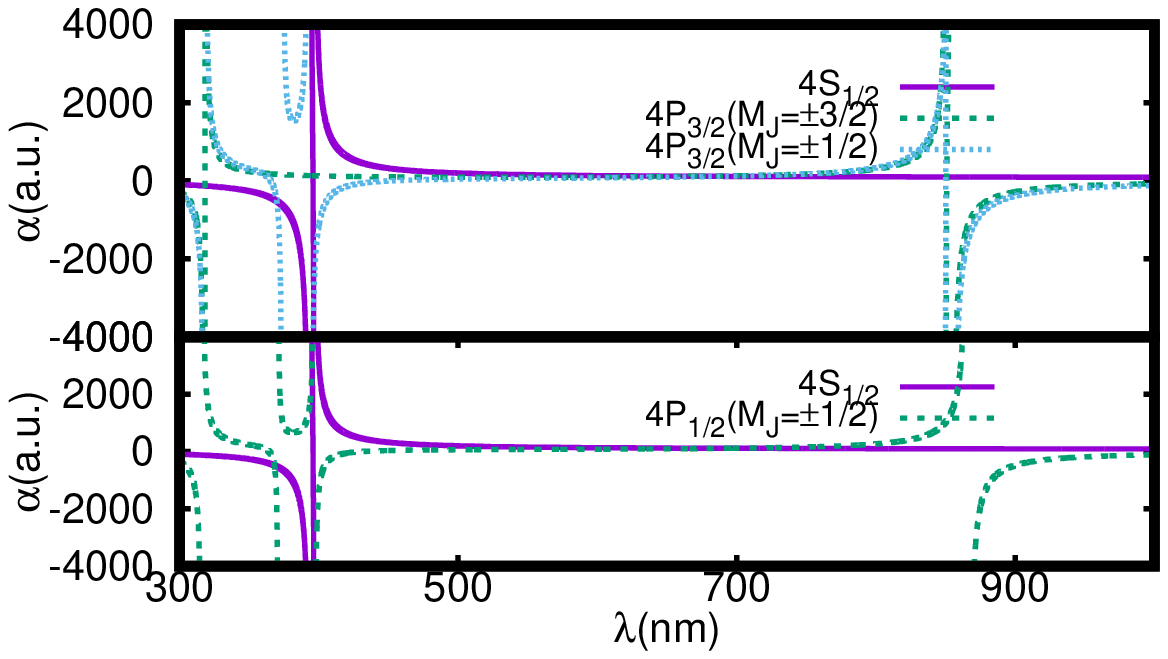} & \includegraphics[height=8cm,width=8.5cm]{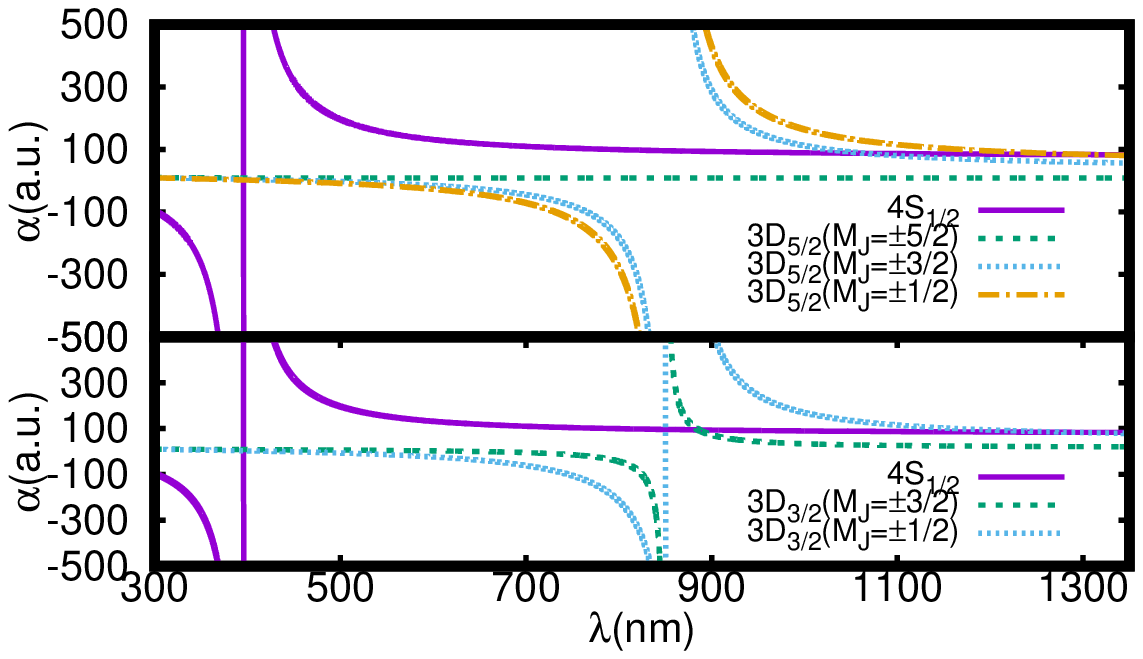} \\
    (a) & (b) \\
\end{tabular}
    \caption{Plots depicting dynamic E1 polarizabilities of (a) the ground $4S_{1/2}$ and excited $4P_{3/2,1/2}$ states, and (b) the ground $4S_{1/2}$ 
    and excited $3D_{5/2,3/2}$ states in Ca$^+$ for the linearly polarized light. The  crossings of polarizability curves  between two resonances correspond to the 
    magic wavelengths.}  \label{camagic}   
\end{figure}

Table \ref{magicca} enlists our evaluated magic wavelengths for the  $4S_{1/2}-4P_{3/2,1/2}$ and $4S_{1/2}-3D_{5/2,3/2}$ transitions of Ca$^+$, 
which are graphically presented in Figs. \ref{camagic}(a) and \ref{camagic}(b), respectively. Our results are compared with the available theoretical
relativistic structure model calculations of Tang et.al.~\cite{tang13}, RCICP 
calculations by Jiang et al.~\cite{jiang2017} and experimental results of Liu et al.~\cite{liu2015}. A total of three magic wavelengths are located 
for the $4S_{1/2}-4P_{1/2}$ transition at 368.08$\pm$0.04 nm, 395.18$\pm$0.02 nm and 693$\pm$2 nm between the $4P_{1/2}-4D_{3/2}$, $4P_{1/2}-5S$, $4P_{1/2}-4S$ and 
$4P_{1/2}-3D_{3/2}$ resonant transitions. As can be clearly noticed that the our results are in excellent agreement with the results given by Tang 
et al.~\cite{tang13} and Jiang et al.~\cite{jiang2017} for magic wavelengths at 368.08$\pm$0.04 nm and 395.18$\pm$0.02 nm.  However, 693$\pm$2 nm shows a slight difference 
with Tang et al.~\cite{tang13} whereas, this value is in perfect agreement and well within the uncertainty bars with the result provided by Jiang
et al.~\cite{jiang2017}. Similarly, many $\lambda_{\rm{magic}}$ values are noticed for the $4S_{1/2}-4P_{3/2}$ transition. Among these the magic 
wavelengths at 369.75$\pm$0.05 nm, 395.77$\pm$0.02 nm, 396.23$\pm$0.01 nm are in very good accord with the results from Ref.~\cite{tang13} as well as Ref.~\cite{jiang2017}. 
The $\lambda_{\rm{magic}}$ at 674$\pm$4 nm and 690$\pm$3 nm are also in perfect accord with the calculations by Jiang et al.~\cite{jiang2017} within error bars 
but our values are slightly red detuned as compared to the values quoted in Ref.~\cite{tang13}. Our results indicate that $\lambda_{\rm{magic}}$s at 
395.77$\pm$0.02 nm and 396.23$\pm$0.01 nm are within the fine-structure splittings of the $4P$ states, which are supported by the experimental measurements by Liu 
et al. ~\cite{liu2015} with magic wavelengths at 395.7992$\pm$0.0007 and 395.7990$\pm$0.0007 nm respectively. However, Ref.~\cite{tang13} have missed the magic 
wavelengths at 850.117$\pm$0.006 nm and 850.92$\pm$0.02 nm between the resonance $4P_{3/2}-3D_{3/2}$ and $4P_{3/2}-3D_{5/2}$ whereas these values from our calculations 
match very well with the values of Ref.~\cite{jiang2017}. It is clear that both for the $4S_{1/2}-3D_{3/2}$ and $4S_{1/2}-3D_{5/2}$ transitions, 
there is a single crossing of polarizability curves at 395.79$\pm$0.01 nm for all the magnetic sublevels.  But, the values of polarizabilities are extremely 
small due to which this magic wavelength is not a good choice for experiments. The other $\lambda_{\rm{magic}}$s at 1288$\pm$3 nm, 1065$\pm$3 nm and 1312$\pm$2 nm are
blue detuned compared to the results presented in Ref.~\cite{tang13} but are in good agreement within uncertainty limits with the values in 
Ref.~\cite{jiang2017}. The large positive value of polarizability which is 2935 a.u., can be noticed for the $\lambda_{\rm{magic}}$ at 395.18$\pm$0.02 nm for the 
$4S_{1/2}-4P_{1/2}$ transition making them a good choice for red detuned traps.

\begin{figure}[t]
    \centering
\begin{tabular}{cc}.
      \includegraphics[height=8cm,width=8.5cm]{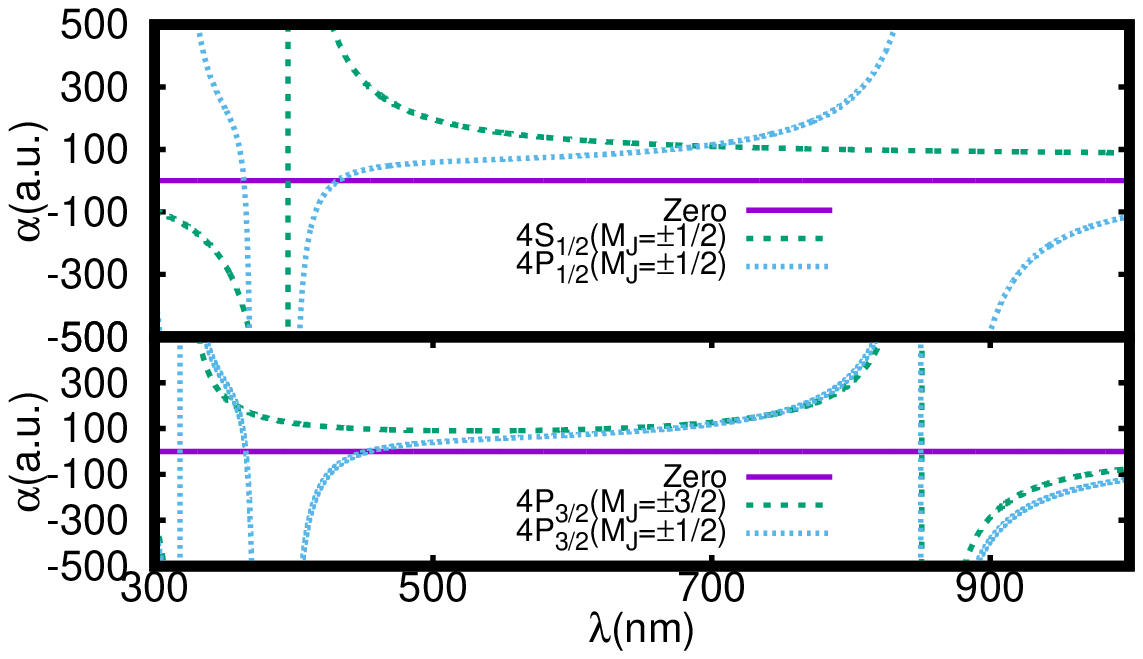} & \includegraphics[height=8cm,width=8.5cm]{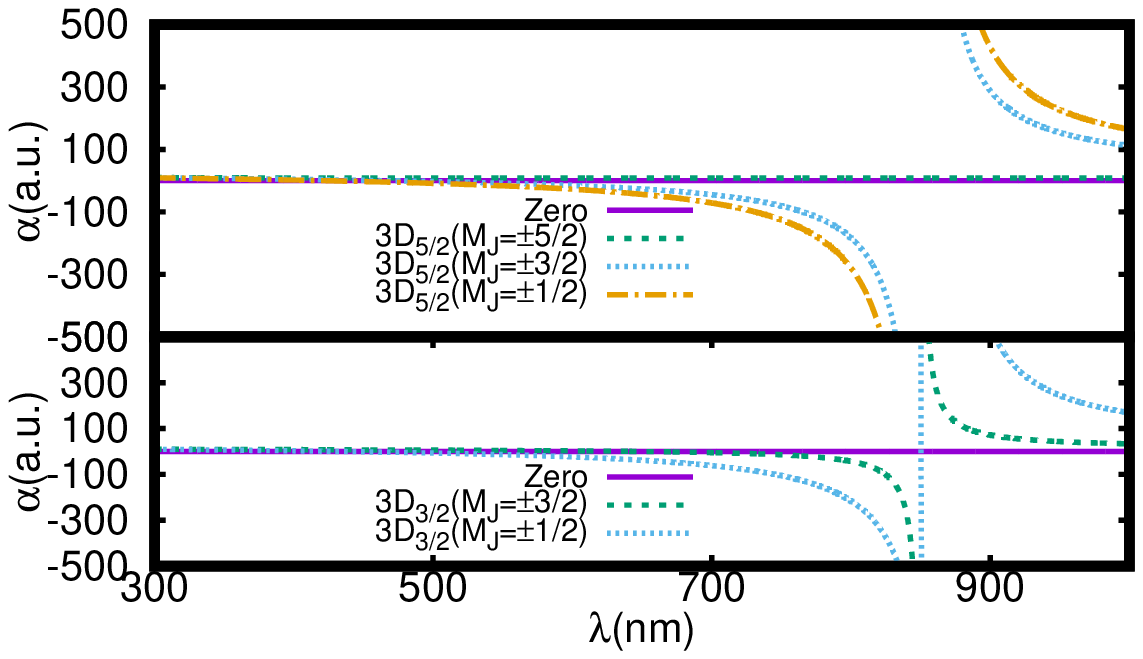}\\
    (a) & (b) \\
\end{tabular}
    \caption{Plots for the dynamic E1 polarizabilities of (a) the ground $4S_{1/2}$ and $4P_{3/2,1/2}$ states, and (b) the $3D_{5/2,3/2}$ states 
    in Ca$^+$ for the linearly polarized light. The crossings of the dynamic dipole polarizability curves with zero   depict
    the tune-out wavelengths.}  \label{catune}  
\end{figure}

Table \ref{tuneout} summarizes the tune-out wavelengths for $nS_{1/2}$, $nP_{3/2,1/2}$ and $(n-1)D_{5/2,3/2}$ states of Ca$^+$ with $n=4$. The 
graphical representation of these tune-out wavelengths can be seen from Figs. \ref{catune}(a) and \ref{catune}(b). The tune-out wavelengths for the 
$4S_{1/2}$ and $4P_{3/2,1/2}$ can be seen in Fig. \ref{catune}(a). The polarizability for the $4S_{1/2}$ state vanishes at 395.80$\pm0.02$ nm which is
in the visible region and lies very adjacent to the $4P_{1/2}-4S_{1/2}$ resonance. The $\lambda_{\rm{T}}$s at 364.13$\pm0.07$ nm and 
431.60$\pm0.85$ nm for the $4P_{1/2}$ state along with $\lambda_{\rm{T}}$s at 365.46$\pm0.07$ nm and 453$\pm1$ nm for the $4P_{3/2}$ state occur away 
from the resonant transitions. However, 850.120$\pm0.009$ nm and 850.900$\pm0.009$ nm are very close to the $4P_{3/2}-3D_{3/2}$ resonance and are in 
infrared region. In a similar way, Fig. \ref{catune}(b) exhibits that the tune-out wavelength at 850.36$\pm0.02$ nm for the $3D_{3/2}$ state is 
adjacent to the $4P_{3/2}-3D_{3/2}$ resonance and is in infrared region. Whereas other $\lambda_{\rm{T}}$s at 433$\pm7$ nm and 647$\pm7$ nm for the 
$3D_{3/2}$ state and at 423$\pm9$ nm and 468$\pm8$ nm for the $3D_{5/2}$ state fall well within the visible range of electromagnetic spectra.

\subsubsection{Sr$^+$}

\begin{figure}[t]
    \centering
\begin{tabular}{cc}.
      \includegraphics[height=8cm,width=8.5cm]{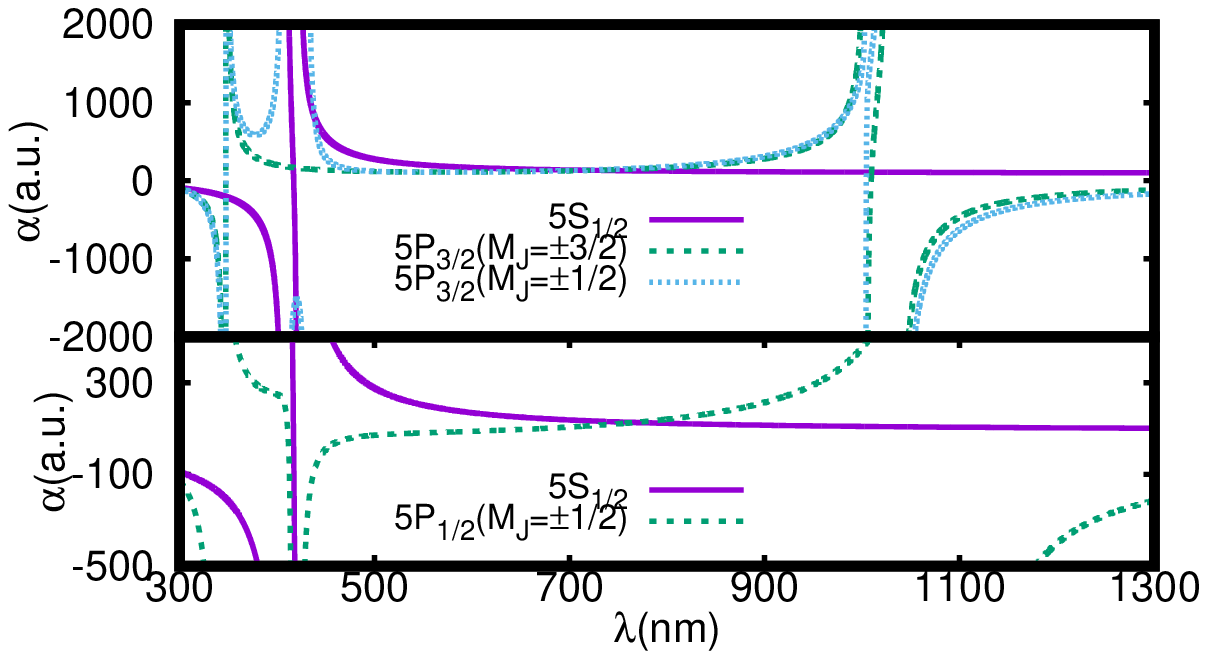} & \includegraphics[height=8cm,width=8.5cm]{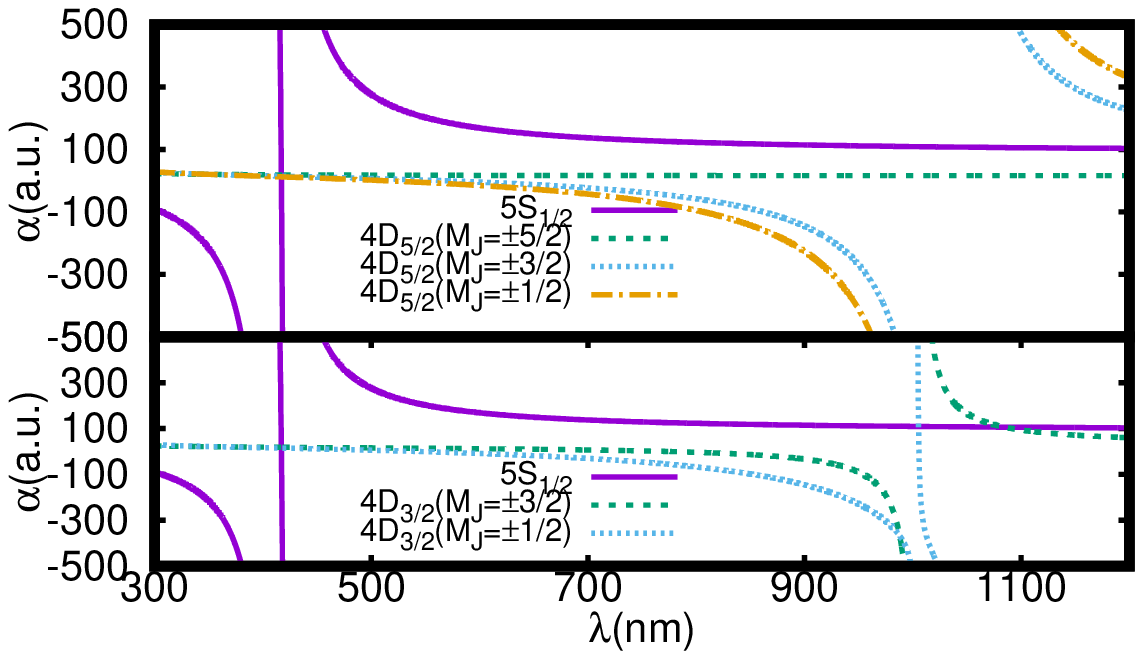} \\
    (a) & (b) \\
\end{tabular}
    \caption{Plots depicting dynamic E1 polarizabilities of (a) the ground $5S_{1/2}$ and excited $5P_{3/2,1/2}$ states, and (b) the ground $5S_{1/2}$ 
    and excited $4D_{5/2,3/2}$ states in Sr$^+$ for the linearly polarized light. The crossings of polarizability curves  between two resonances correspond to the 
    magic wavelengths.}  \label{srmagic}   
\end{figure}

The calculated $\lambda_{\rm{magic}}$s for the $5S-5P_{1/2}$ and $5S-4D_{5/2,3/2}$ transitions of Sr$^+$ are presented in Table \ref{magicsr} 
by locating at the crossings of the polarizability curves for the considered states as shown in Figs. \ref{srmagic}(a) and \ref{srmagic}(b). The 
polarizability values at the respective magic wavelengths along with different resonant lines are also listed in the same table. The only magic 
wavelength available for the $5S-5P_{1/2}$ transition at 767$\pm$2 nm is in good agreement with the value reported by Jiang et al.~\cite{jiang2016}, who 
have used relativistic semi-empirical-core-potential approach using Laguerre and Slater spinors to determine it. However, a series of magic 
wavelengths for the $5S-5P_{3/2}$ transition are located between various resonances. The $\lambda_{\rm{magic}}$=438.5$\pm$0.2 nm matches excellently with 
the value given by~\cite{jiang2016}. The values of polarizabilities for all the $\lambda_{\rm{magic}}$ are very small except for the magic wavelength 
at 419.31$\pm$0.04 nm, which has a very large negative value of polarizability $-1526$ a.u.. Hence, it is very useful for trapping Sr$^+$ ion using blue detuned traps.
Some other $\lambda_{\rm{magic}}$s at 713$\pm$2 nm and 722$\pm$2 nm arise due to gradual increase in the $5S_{1/2}$ state polarizability and the gradual decrease 
in the $5P_{3/2}$ state polarizability when the wavelength approaches the $5S-5P_{j}$ resonant wavelength. Similarly, two $\lambda_{\rm{magic}}s$ at 
1004.47$\pm$0.05 nm and $\lambda_{\rm{magic}}$=1009.8$\pm$0.1 nm lie in the infrared region and they are recommended for the red detuned traps. All of these 
$\lambda_{\rm{magic}}$ values,  at 713$\pm$2 nm, 722$\pm$2 nm, 1004.47$\pm$0.05 nm and 1009.8$\pm$0.1 nm, are in close agreement with the values reported in 
Ref.~\cite{jiang2016}. For the $\lambda_{\rm{magic}}$s at 416.99$\pm$0.06 nm and 417.01$\pm$0.06 nm for the $5S-4D$ transitions, the associated polarizability values 
are extremely small, so are of limited experimental use. $\lambda_{\rm{magic}}$ at 1085 nm, which is in good accord with the values calculated in 
Ref.~\cite{jiang2016}, is in vicinity of the $4D_{3/2}-5P_{3/2}$ resonant transition. 

\begin{figure}[t]
    \centering
\begin{tabular}{cc}.
      \includegraphics[height=8cm,width=8.5cm]{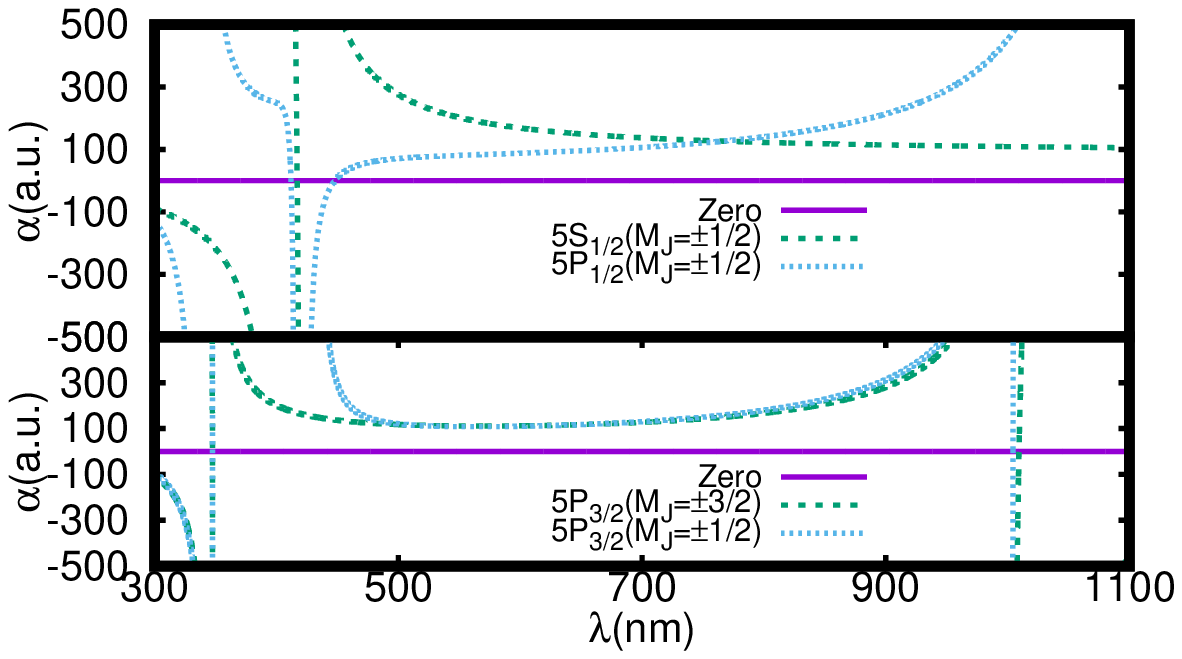} & \includegraphics[height=8cm,width=8.5cm]{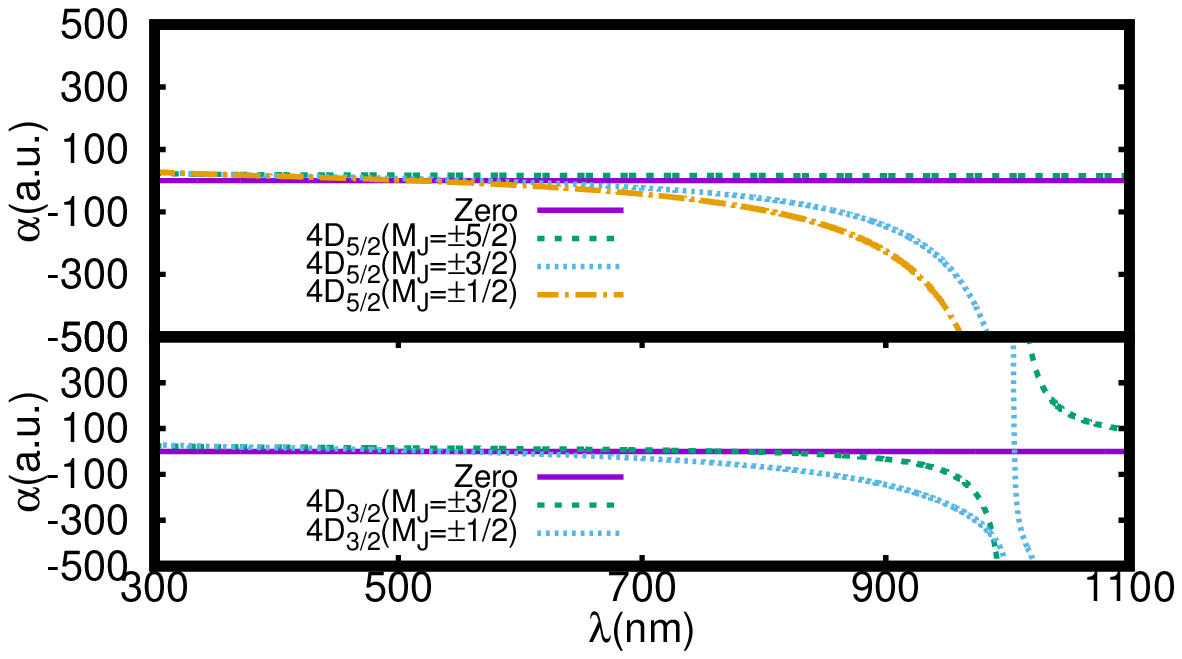}\\
    (a) & (b) \\
\end{tabular}
    \caption{Plots for the dynamic E1 polarizabilities of (a) the ground $5S_{1/2}$ and $5P_{3/2,1/2}$ states, and (b) the $4D_{5/2,3/2}$ states 
    in Sr$^+$ for the linearly polarized light. The crossings of the dynamic dipole polarizability curves with zero  depict
    the tune-out wavelengths.}  \label{srtune}  
\end{figure}

We have also provided the tune-out wavelengths for the $nS_{1/2}$, $nP_{3/2,1/2}$ and $(n-1)D_{5/2,3/2}$ states of Sr$^+$ with $n=5$ by plotting 
dynamic polarizabilities in the Figs. \ref{srtune}(a) and \ref{srtune}(b). These values are also tabulated in Table \ref{tuneout}. The graphical 
representation of the tune-out wavelengths for the $5S_{1/2}$ and $5P_{3/2,1/2}$ states is given in Fig. \ref{srtune}(a). It can be inferred from the 
figure that for the  $5P_{3/2,1/2}$ states two $\lambda_{\rm{T}}$s at 1004.43$\pm0.05$ nm  and 1009.37$\pm0.12$ nm lie in infrared region whereas 
other tune-out wavelengths lie in the visible region. Similarly, it can be noticed from Fig. \ref{srtune}(b) that all the $\lambda_{\rm{T}}$s for the 
$4D_{3/2}$ and $4D_{5/2}$ states are in visible region except 1005.67$\pm0.16$ nm which is in the infrared region.

\subsubsection{Ba$^+$}

\begin{figure}[t]
    \centering
\begin{tabular}{cc}.
      \includegraphics[height=8cm,width=8.5cm]{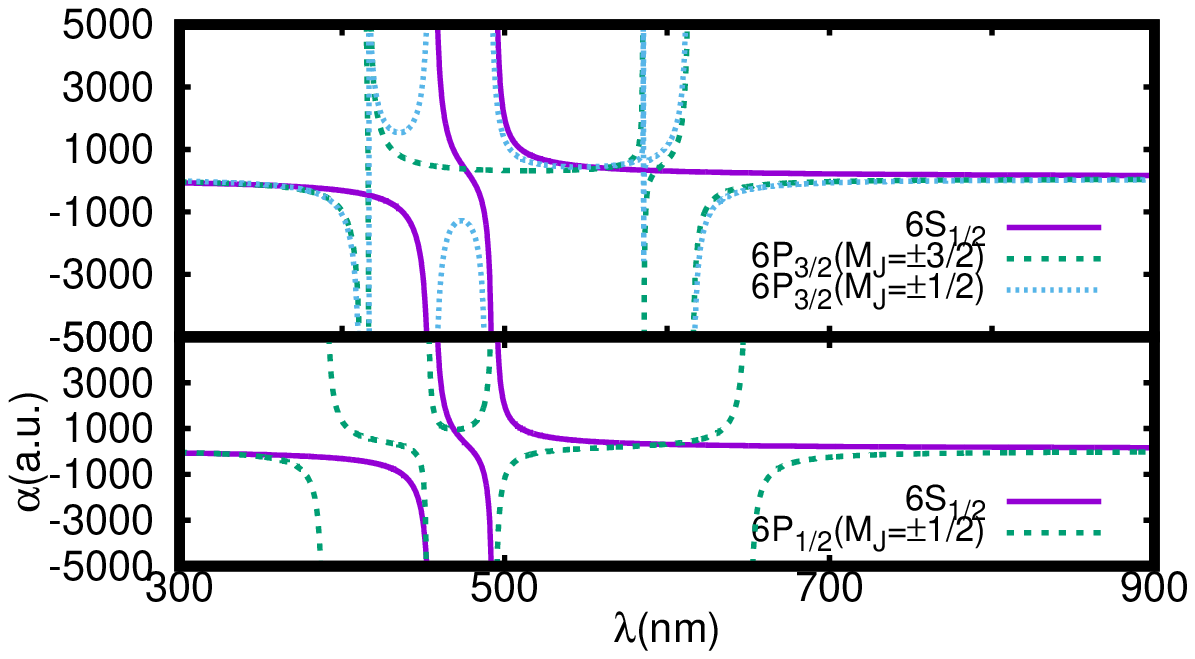} & \includegraphics[height=8cm,width=8.5cm]{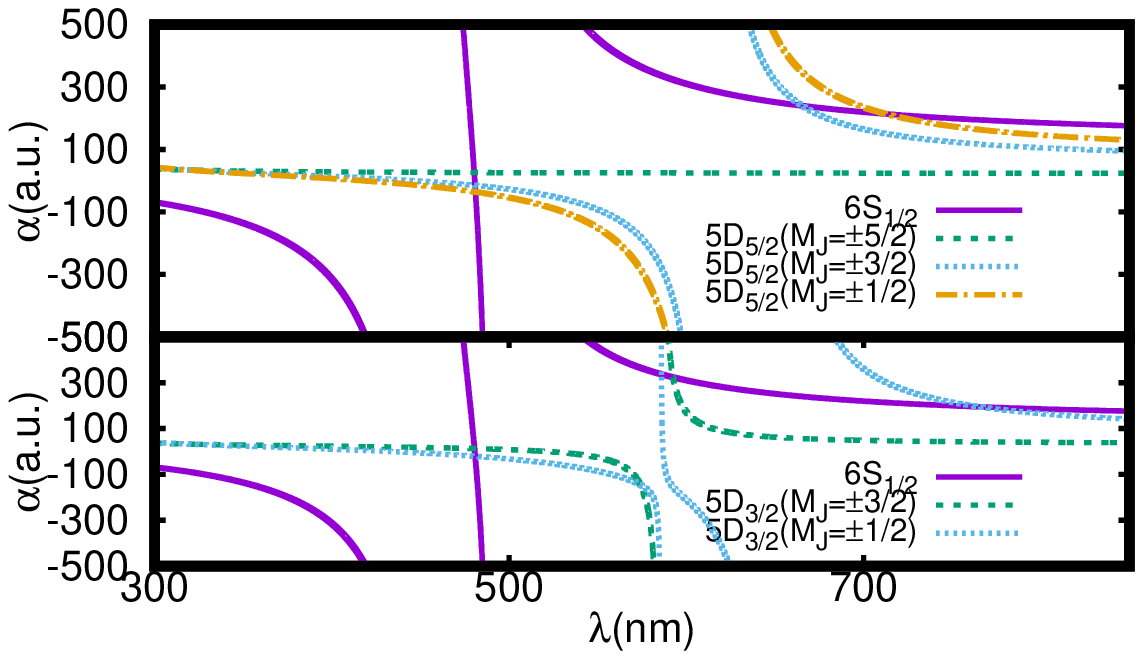} \\
    (a) & (b) \\
\end{tabular}
    \caption{Plots depicting dynamic E1 polarizabilities of (a) the ground $6S_{1/2}$ and excited $6P_{3/2,1/2}$ states, and (b) the ground $6S_{1/2}$ 
    and excited $5D_{5/2,3/2}$ states in Ba$^+$ for the linearly polarized light. The crossings of polarizability curves  between two resonances correspond to the 
    magic wavelengths.}  \label{bamagic}   
\end{figure}

The magic wavelengths for the Ba$^+$ presented in Figs. \ref{bamagic}(a) and \ref{bamagic}(b) are summarized in Table \ref{magicba}. Four 
$\lambda_{\rm{magic}}$ are found for the $6S_{1/2}-6P_{1/2}$ transition in Ba$^+$ at 323$\pm$2 nm, 451.72$\pm$0.04 nm, 468.9$\pm$0.3 nm and 599$\pm$1 nm with 
$\lambda_{\rm{magic}}$ at 451.72$\pm$0.04 nm yielding a very large negative value of polarizability $-4557$ a.u. which makes it a useful wavelength for 
trapping Ba$^+$. Also, the $\lambda_{\rm{magic}}$ 468.9$\pm$0.3 carries significant value of polarizability. We also identify at least five 
$\lambda_{\rm{magic}}$ for the $6S_{1/2}-6P_{3/2}$ transition between various resonant transitions. Out of these, magic wavelengths around 416 nm, 
552$\pm$2 nm ans 561$\pm$1 nm hold reasonably large values of polarizability. Similarly, several $\lambda_{\rm{magic}}$s are also located for the $6S-5D_{3/2}$ 
and $6S-5D_{5/2}$ transitions in the wavelength range $300-800$ nm. Among these, the wavelengths around 480 nm are of negligible use due to very 
small values of associated polarizability. Some other $\lambda_{\rm{magic}}$s at 585.99$\pm$0.05 nm, 592.5$\pm$0.2 nm, 766$\pm$4 nm, 665$\pm$2 nm and 713$\pm$3 nm are also presented with comparatively good values of polarizabilities. 

\begin{figure}[t]
    \centering
\begin{tabular}{cc}.
      \includegraphics[height=8cm,width=8.5cm]{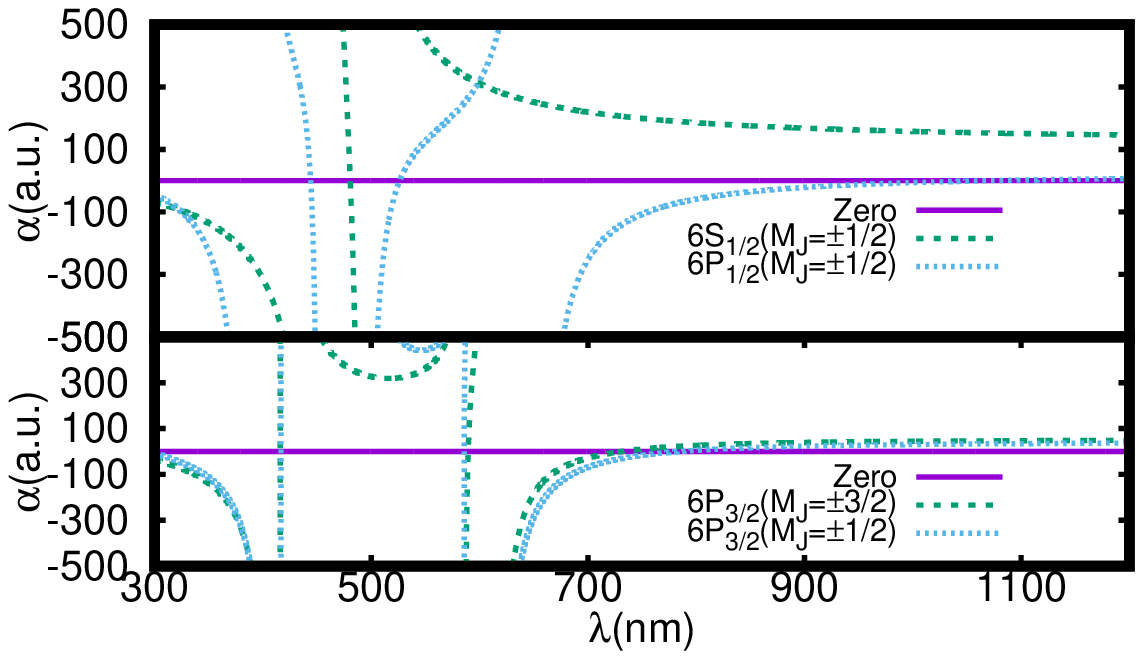} & \includegraphics[height=8cm,width=8.5cm]{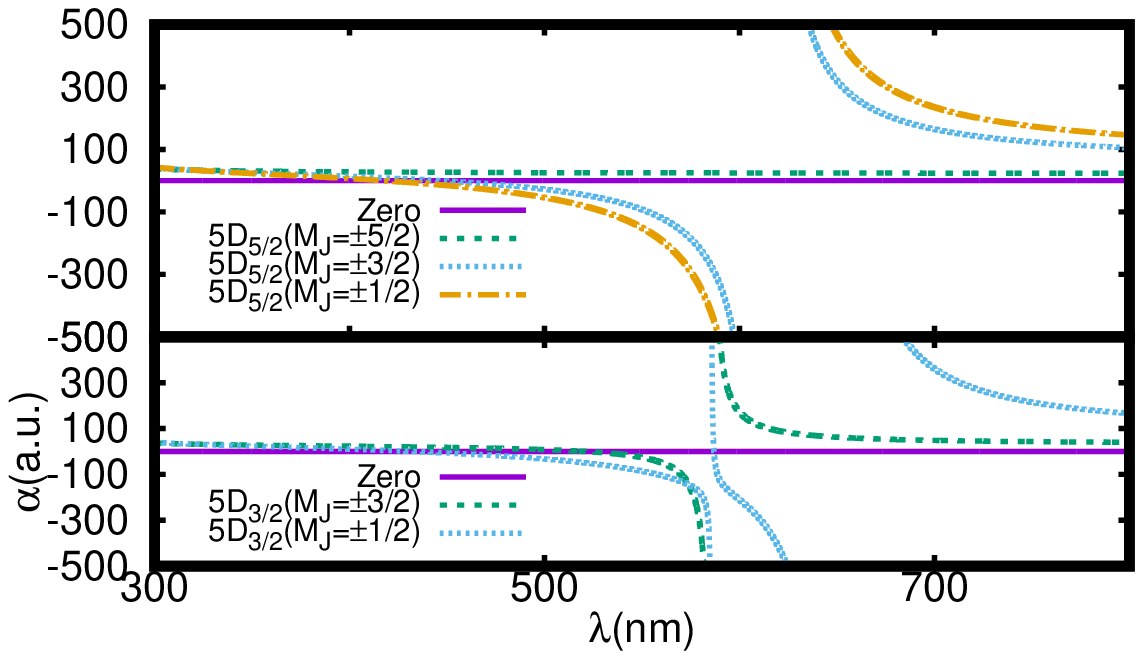}\\
    (a) & (b) \\
\end{tabular}
    \caption{Plots for the dynamic E1 polarizabilities of (a) the ground $6S_{1/2}$ and $6P_{3/2,1/2}$ states, and (b) the $5D_{5/2,3/2}$ states 
    in Ba$^+$ for the linearly polarized light. The crossings of the dynamic dipole polarizability curves with zero   depict
    the tune-out wavelengths.}  \label{batune}  
\end{figure}

Along with this, the tune-out wavelengths are shown graphically in Figs. \ref{batune}(a) and \ref{batune}(b) for the $nS_{1/2}$, $nP_{3/2,1/2}$ and 
$(n-1)D_{5/2,3/2}$ states of Ba$^+$ with $n=6$ and they are also tabulated in Table \ref{tuneout}. It can be inferred from Fig. \ref{batune}(a) that 
the only $\lambda_{\rm{T}}$ at 480.63$\pm$0.24 nm for $6S_{1/2}$ occurs between the $6P_{1/2}-6S_{1/2}$ and $6P_{1/2}-7S_{1/2}$ resonances and is in 
the visible region. Along with it, we find three $\lambda_{\rm{T}}$s for the $6P_{1/2}$ state at 444.06$\pm$0.17 nm, 526.29$\pm$0.80 nm in visible 
region and 1071$\pm$6 nm in infrared region. For the $6P_{3/2}$ state, $\lambda_{\rm{T}}$s at 585.88$\pm$0.04 nm and 589.50$\pm$0.07 nm lie very close 
to the $6P_{3/2}-5D_{3/2}$ resonance. Two $\lambda_{\rm{T}}$s lie in the vicinity of the $6P_{3/2}-6D_{3/2}$ resonance at 416.65$\pm$0.01 nm and 
416.00$\pm$0.01 nm. In the visible region, there are two tune-out wavelengths at 731$\pm$5 nm and 785$\pm$5 nm for this state. In a similar way, 
Fig. \ref{batune}(b) presents the $\lambda_{\rm{T}}$s for the $5D_{3/2}$ and $5D_{5/2}$ states. A total of 5 tune-out wavelengths were found for these
two states in the considered wavelength range, and all of them are in the visible region.

\subsection{Dynamic quadrupole polarizabilities}

\begin{figure}[t]
    \centering
\begin{tabular}{cc}.
      \includegraphics[height=8cm,width=8.5cm]{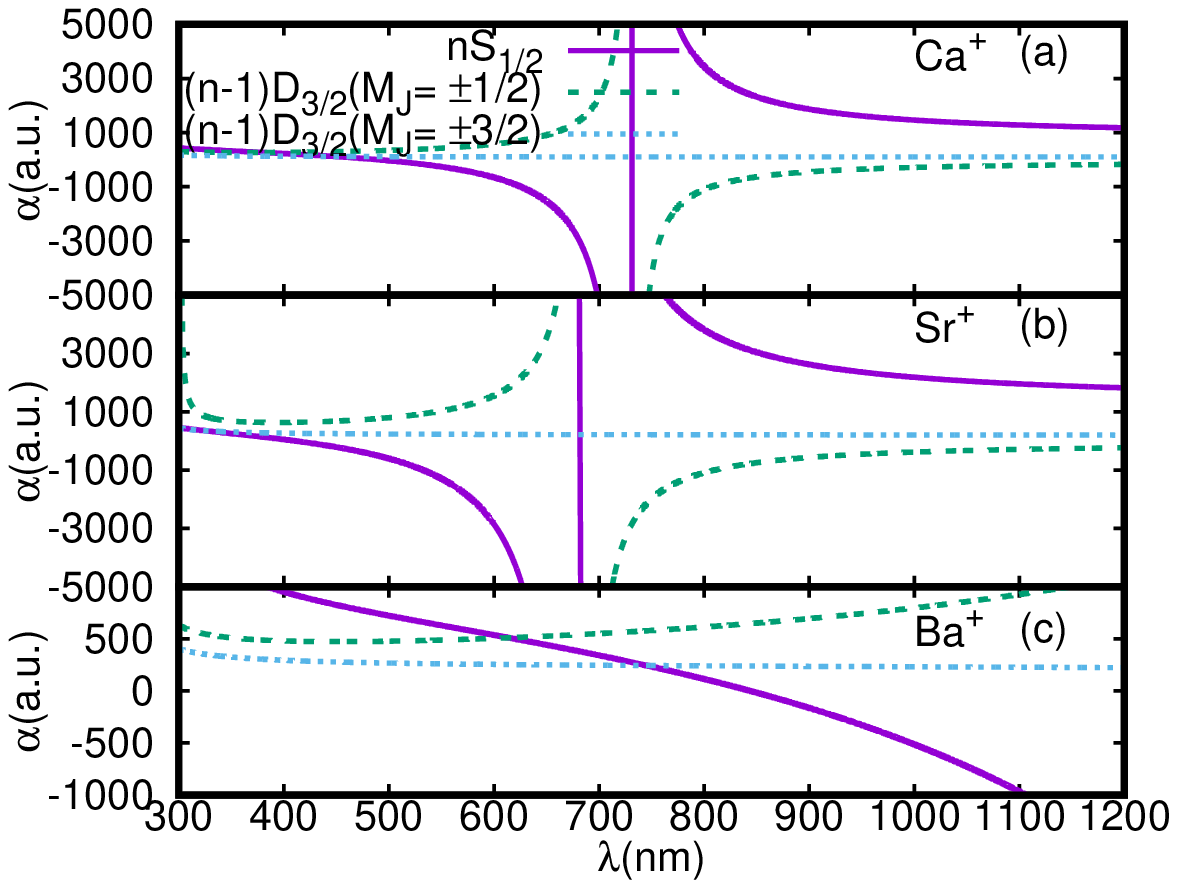} & \includegraphics[height=8cm,width=8.5cm]{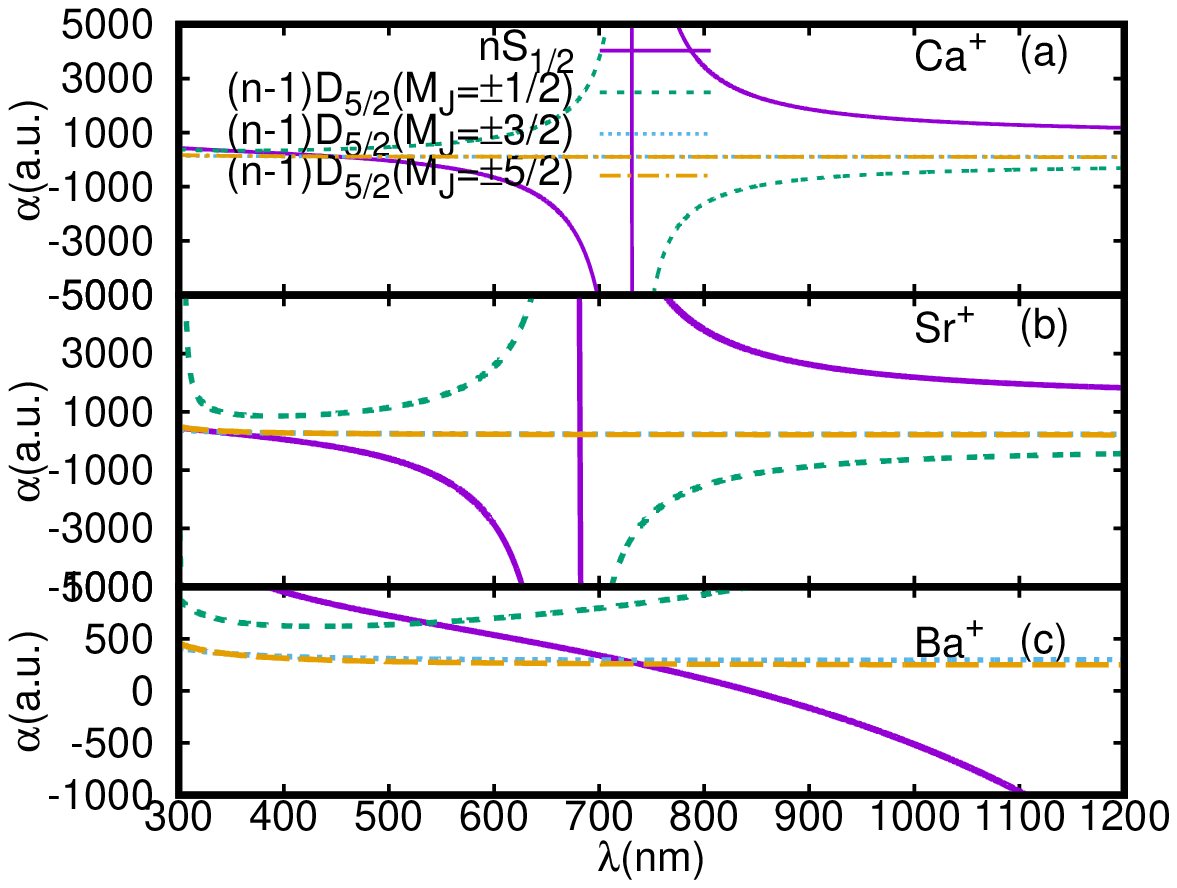} \\
    (a) & (b) \\
\end{tabular}
    \caption{Plots depicting dynamic E2 polarizabilities of (a) ground and metastable $(n-1)D_{3/2}$ states in Ca$^+$(n=4), Sr$^+$(n=5) and Ba$^+$(n=6) (b) of ground and the metastable $(n-1)D_{5/2}$ states in Ca$^+$(n=4), Sr$^+$(n=5) and Ba$^+$(n=6) for wavelength range $300 - 1300$ nm for the linearly polarized light.}  \label{quad3}   
\end{figure}

Here, we discuss the dynamic quadrupole polarizabilities of the ground $nS_{1/2}$ and metastable $(n-1)D_{5/2,3/2}$ states of Ca$^+$, Sr$^+$ and 
Ba$^+$ ions for the principal quantum number $n$ of the respective ion using Eqs. (\ref{eq-quad})-(\ref{w3}). It should be noted that for the 
$nS_{1/2}$ states only the scalar part contributes, whereas for the $(n-1)D_{3/2}$ and the $(n-1)D_{5/2}$ states, both the scalar and tensor 
components contribute to the total values of quadrupole polarizabilities.

The dynamic quadrupole polarizabilities of Ca$^+$, Sr$^+$ and Ba$^+$ for the $nS_{1/2}$ and $(n-1)D_{3/2}$ states are shown in Fig.~\ref{quad3}(a) for 
the wavelength range from 300 nm to 1300 nm. In the case of the $(n-1)D_{3/2}$ states, $\alpha^q_{n4}$ is zero. The quadrupole polarizability for 
these states need to be determined separately for the cases with $M_J=\pm1/2$ and $M_J=\pm3/2$ owing to the presence of the tensor contributions to 
the total quadrupole polarizabilities of the $D_{3/2}$ states. As shown in Fig.~\ref{quad3}(a), in the considered wavelength range, the $4S_{1/2}$ state 
quadrupole polarizability for Ca$^+$ has two E2 resonances in the $3D_{3/2}-4S_{1/2}$ and $3D_{5/2}-4S_{1/2}$ transitions at 732.389 and 729.147 nm, 
respectively. Thus, the quadrupole polarizability for the $4S_{1/2}$ state diverges at these two wavelength values. The $3D_{3/2}-4S_{1/2}$ resonant 
transition also contributes to the quadrupole polarizability of the $3D_{3/2}(M_J=\pm1/2)$ level. As a result, quadrupole polarizabilities for the 
$4S_{1/2}$ and $3D_{3/2}(M_J=\pm1/2)$ levels are large but they have opposite signs in the vicinity of this resonance. However, for the 
$3D_{3/2}(M_J=\pm3/2)$ level, when $\alpha^q_n=\alpha^q_{n0}+\alpha^q_{n2}$, the $3D_{3/2}-n'S_{1/2}$ transitions, with $n'$ referring to higher 
excited quantum number, do not contribute to the total quadrupole polarizability of the $3D_{3/2}$ state owing to the exact cancellation of the 
scalar $\alpha_{n0}^q$ and tensor $\alpha_{n2}^q$ contributions from the $3D_{3/2}-n'S_{1/2}$ transitions as can be understood using 
Eq. (\ref{alphat}). In this case, the $W_{n,k}^{q(2)}$ factor of Eq. (\ref{w2}) is exactly equal to $-W_n^{q(0)}$ of Eq. (\ref{w1}) leading 
to exact cancellation of the contributions from the $3D_{3/2}-n'S_{1/2}$ transitions, and contributions to the total polarizability comes only from 
the $3D_{3/2}-n'D_{5/2,3/2}$ and $3D_{3/2}-n'G_{9/2,7/2}$ transitions, which do not cancel out and are small. As a result, there is no resonance for
the $M_J=\pm 3/2$ case at the wavelength corresponding to the $3D_{3/2}-4S_{1/2}$ transition leading to a straight quadrupole polarizability curve 
for this state (shown in blue dotted curve in Fig.~\ref{quad3}(a)). A similar trend in the quadrupole polarizability for Sr$^+$ ion is expected with two 
E2 resonances appearing in ground state polarizability between the 300-1300 nm range, the $4D_{3/2}-5S_{1/2}$ transition at 686.8171 nm and the 
$4D_{5/2}-5S_{1/2}$ transition at 673.8392 nm. The quadrupole polarizability for the $4D_{3/2}(M_J=\pm1/2)$ level is small in this range except  
in the close vicinity of the $4D_{3/2}-5S_{1/2}$ transition. For Sr$^+$ ion as well, the $4D_{3/2}-5S_{1/2}$ transition does not contribute to the 
quadrupole polarizability of the $4D_{3/2}(M_J=\pm3/2)$ level. The case for the quadrupole polarizability of Ba$^+$ is different. We did not find 
any E2 resonant transitions for the $6S_{1/2}$ and $5D_{3/2}$ states in the considered wavelength range. The corresponding quadrupole polarizability 
is generally not as high as it is for the other two ions in the considered wavelength range.

In Fig. \ref{quad3}(b), the dynamic quadrupole polarizabilities of the $nS_{1/2}$ and $(n-1)D_{5/2}$ states, with principal quantum number $n$, of the considered ions 
are presented. In this case, the rank four component of the tensor quadrupole polarizabilities are also involved. The polarizability curves for 
different values of $M_J$ (1/2, 3/2 and 5/2) are separately shown in the above figure. It can be noticed that the polarizability curves for 
$J=5/2$ exhibit same trend as that for $J=3/2$. The quadrupole polarizability curve for $M_J=\pm 5/2$ almost overlaps with the polarizability curve 
for $M_J=\pm 3/2$ in the Ca$^+$, Sr$^+$ and Ba$^+$ ions. There is no E2 resonance line found for the Ba$^+$ ion in the considered wavelength range 
and it does not attain a very large value anywhere in this range.

\begin{longtable}{@{\extracolsep\fill}rllllllll@{}}
\caption{\label{matrixmge2} Reduced E2 matrix elements (in a.u.) of Mg$^+$ from the DHF, SD and SDpT methods, whereas the label ``sc'' indicates the scaled values. The second last and the last columns list the recommended ${\rm{\textit{O}}}^{\rm{{Final}}}$ values of the matrix elements and their absolute uncertainties, respectively.}\\
\hline
\hline
Transition   & {${\rm{\textit{O}}}^{\rm{DHF}}$} &  {${\rm{\textit{O}}}^{\rm{SD}}$} & {${\rm{\textit{O}}}^{\rm{SD}}_{\rm{sc}}$} &  {${\rm{\textit{O}}}^{\rm{SDpT}}$} & {${\rm{\textit{O}}}^{\rm{SDpT}}_{\rm{sc}}$ }& {${\rm{\textit{O}}}^{\rm{Final}}$} & {${\rm{\delta}}$} \\
\hline
 $3P_{3/2}-3P_{1/2}$&12.721   & 12.309   &  12.312   &  12.309 & 12.311  & 12.312 &  0.003   \\
 $3D_{5/2}-3S_{1/2}$&12.516 & 12.092 & 12.094 & 12.093 & 12.094 &12.094 &0.001\\
 $3D_{5/2}-4S_{1/2}$&37.631 & 36.804 & 36.800 & 36.808 & 36.799 &36.800 &0.008\\
 $3D_{3/2}-3S_{1/2}$&10.219& 9.873  & 9.875 & 9.874 & 9.876 & 9.875 &0.001\\
 $3D_{3/2}-4S_{1/2}$&30.727 & 30.051 & 30.047 & 30.055 & 30.046 &30.047 &0.008\\
 $3D_{3/2}-3D_{5/2}$&17.732 & 17.297 & 17.294 & 17.299 & 17.293 &17.294 &0.005\\
\end{longtable}

\begin{longtable}{@{\extracolsep\fill}rlllllll@{}}
\caption{\label{matrixcae2} Reduced E2 matrix elements (in a.u.) of Ca$^+$ from the DHF, SD and SDpT methods, whereas the label ``sc'' indicates the scaled values. The second last and the last columns list the recommended ${\rm{\textit{O}}}^{\rm{{Final}}}$ values of the matrix elements and their absolute uncertainties, respectively.}\\
\hline
\hline
Transition   & {${\rm{\textit{O}}}^{\rm{DHF}}$} & {${\rm{\textit{O}}}^{\rm{SD}}$} & {${\rm{\textit{O}}}^{\rm{SD}}_{\rm{sc}}$} &  {${\rm{\textit{O}}}^{\rm{SDpT}}$} & {${\rm{\textit{O}}}^{\rm{SDpT}}_{\rm{sc}}$ }& {${\rm{\textit{O}}}^{\rm{Final}}$} & {${\rm{\delta}}$} \\
\hline

 $3D_{3/2}-4S_{1/2}$ &  9.767 &  7.788 &  7.945 &  7.971 &  7.907 & 7.94 & 0.04\\
 $3D_{5/2}-4S_{1/2}$ & 11.978 &  9.561 &  9.750 &  9.785 &  9.704 & 9.75  &0.04 \\
 $3D_{5/2}-3D_{3/2}$ &  5.018&  3.688&  3.781&  3.794&  3.758& 3.78 &0.02\\
 $4P_{3/2}-4P_{1/2}$ & 20.820 & 19.259  & 19.366   &  19.414  &  19.360  &  19.36 & 0.05 \\
 $5S_{1/2}-3D_{3/2}$ &  6.050 &  3.736 &  3.955 &  3.933 &  3.864 & 3.95 &0.09 \\
 $5S_{1/2}-3D_{5/2}$ &  7.427 &  4.600 &  4.863 &  4.841 &  4.752 &  4.86 &0.11\\
\end{longtable}

\begin{longtable}{@{\extracolsep\fill}rlllllll@{}}
\caption{\label{matrixsre2} Reduced E2 matrix elements (in a.u.) of Sr$^+$ from the DHF, SD and SDpT methods, whereas the label ``sc'' indicates the scaled values. The second last and the last columns list the recommended ${\rm{\textit{O}}}^{\rm{{Final}}}$ values of the matrix elements and their absolute uncertainties, respectively.}\\
\hline
\hline
Transition  & {${\rm{\textit{O}}}^{\rm{DHF}}$} & {${\rm{\textit{O}}}^{\rm{SD}}$} & {${\rm{\textit{O}}}^{\rm{SD}}_{\rm{sc}}$} & {${\rm{\textit{O}}}^{\rm{SDpT}}$} & {${\rm{\textit{O}}}^{\rm{SDpT}}_{\rm{sc}}$ }  & {${\rm{\textit{O}}}^{\rm{Final}}$} & {${\rm{\delta}}$}\\
\hline 
 $4D_{3/2}-5S_{1/2}$ & 12.968& 11.009& 11.133& 11.172& 11.100& 11.13 & 0.04\\
 $4D_{5/2}-5S_{1/2}$ & 15.972& 13.601& 13.747& 13.797& 13.706& 13.74 &0.05\\
 $4D_{5/2}-4D_{3/2}$ &  7.260&  5.904&  5.978&  6.001&  5.954& 5.97 &0.02\\
 $5P_{3/2}-5P_{1/2}$ &25.408  & 23.257  & 23.406  & 23.462  & 23.393 & 23.40 &0.05 \\
 $6S_{1/2}-4D_{3/2}$ &  7.969&  5.879&  6.031&  6.041&  5.957&  6.03 &0.07\\
 $6S_{1/2}-4D_{5/2}$ &  9.885&  7.356&  7.531&  7.550&  7.441& 7.53 & 0.09\\
\hline		                      
\end{longtable}

\begin{longtable}{@{\extracolsep\fill}rlllllll@{}}
\caption{\label{matrixbae2} Reduced E2 matrix elements (in a.u.) of Ba$^+$ from the DHF, SD and SDpT methods, whereas the label ``sc'' indicates the scaled values. The second last and the last columns list the recommended ${\rm{\textit{O}}}^{\rm{{Final}}}$ values of the matrix elements and their absolute uncertainties, respectively.}\\
\hline
\hline
Transition  & {${\rm{\textit{O}}}^{\rm{DHF}}$}  & {${\rm{\textit{O}}}^{\rm{SD}}$} & {${\rm{\textit{O}}}^{\rm{SD}}_{\rm{sc}}$} &  {${\rm{\textit{O}}}^{\rm{SDpT}}$} & {${\rm{\textit{O}}}^{\rm{SDpT}}_{\rm{sc}}$ } & {${\rm{\textit{O}}}^{\rm{Final}}$} & {${\rm{\delta}}$} \\
\hline
 $5D_{3/2}-6S_{1/2}$ & 14.763& 12.497& 12.627& 12.740& 12.589& 12.62 &0.11\\
 $5D_{5/2}-6S_{1/2}$ & 18.383& 15.651& 15.803& 15.943& 15.758& 15.80 &0.14\\
 $5D_{5/2}-5D_{3/2}$ &  8.091&  6.649&  6.708&  6.772&  6.686& 6.70  &0.06\\
 $6P_{3/2}-6P_{1/2}$ & 31.319 &  28.079& 28.336& 28.472 & 28.322 & 28.33  & 0.13  \\
 $7S_{1/2}-5D_{3/2}$ &  6.199&  4.657&  4.741&  4.806&  4.650& 4.74 & 0.09\\
 $7S_{1/2}-5D_{5/2}$ &  7.936&  6.075&  6.164&  6.248&  6.056& 6.16 &0.11 \\
\end{longtable}


\vspace{1 cm}
\begin{longtable}{@{\extracolsep\fill}rllllllll@{}}
\caption{\label{parametersmge2}Transition wavelengths ($\lambda$) and electric quadrupole transition properties such as line strengths ($S_{vk}$), 
transition probabilities ($A_{vk}$) and oscillator strengths ($f_{kv}$) along with their uncertainties between transitions involving an upper
energy level (denoted by $v$) and lower energy level (denoted by $k$) of Mg$^+$. Values given in square brackets represent the power of 10.}\\
\hline
\hline
\multicolumn{2}{c}{Levels} & $\lambda$  & $S_{vk}$ & $A_{vk}$ & $f_{kv}$  \\ 
\cline{1-2}
Upper $(v)$ & Lower $(k)$ & (in {\AA}) & (in a.u.) & (in $\rm{s^{-1}}$) &  \\
\hline
\hline
  3$P_{3/2}$&  3$P_{1/2}$&       1092060.718  &1.5151$\pm$0.0007[2] &2.732$\pm$0.001[-11]  &9.769$\pm$0.004[-15]\\
  3$D_{5/2}$&  3$S_{1/2}$&         1398.793&   1.4631$\pm$0.0002[2]& 5.0971$\pm$0.0008[3] &  4.4860$\pm$0.0007[-6]   \\
  3$D_{5/2}$&  4$S_{1/2}$&        59338.729&   1.3540$\pm$0.0005[3]& 3.435$\pm$0.001[-4] & 5.441$\pm$0.002[-10]   \\
  3$D_{3/2}$&  3$S_{1/2}$&         1398.776&   9.752$\pm$0.002[1]& 5.098$\pm$0.001[3]&   2.9911$\pm$0.0006[-6]   \\
  3$D_{3/2}$&  4$S_{1/2}$&        59308.112&   9.028$\pm$0.005[2]& 3.444$\pm$0.002[-4]&  3.633$\pm$0.002[-10]   \\
\end{longtable}

\vspace{1 cm}
\begin{longtable}{@{\extracolsep\fill}rllllllll@{}}
\caption{\label{parameterscae2} Transition wavelengths ($\lambda$) and electric quadrupole transition properties such as line strengths ($S_{vk}$), 
transition probabilities ($A_{vk}$) and oscillator strengths ($f_{kv}$) along with their uncertainties between transitions involving an upper
energy level (denoted by $v$) and lower energy level (denoted by $k$) of Ca$^+$. Values given in square brackets represent the power of 10.}\\
\hline
\hline
\multicolumn{2}{c}{Levels} & $\lambda$  & $S_{vk}$ & $A_{vk}$ & $f_{kv}$  \\ 
\cline{1-2}
Upper $(v)$ & Lower $(k)$ & (in {\AA}) & (in a.u.) & (in $\rm{s^{-1}}$) &  \\
\hline
\hline
$3D_{3/2} $ & $ 4S_{1/2}$   &         7325.905   & 6.31$\pm$0.06[1]   &0.838$\pm$0.008[0]   &1.35$\pm$0.01[-8]   \\
$3D_{5/2} $ & $ 4S_{1/2}$   &         7293.477   & 9.51$\pm$0.07[1]   &0.860$\pm$0.007[0]   &2.05$\pm$0.02[-8]   \\
$3D_{5/2} $ & $ 3D_{3/2}$   &      1647717.911   & 1.43$\pm$0.01[1]   &2.19$\pm$0.02[-13]  &1.34$\pm$0.01[-16] \\
$4P_{3/2} $ & $ 4P_{1/2}$   &         448651.801 & 3.75$\pm$0.02[2]   &5.77$\pm$0.03[-9]   & 3.48$\pm$0.02[-13]  \\
$5S_{1/2} $ & $ 3D_{3/2}$   &         2596.273   & 1.56$\pm$0.07[1]   &7.42$\pm$0.34[1]    &3.75$\pm$0.17[-8]   \\
\hline
\hline

\end{longtable}

\begin{longtable}{@{\extracolsep\fill}rllllllll@{}}
\caption{\label{parameterssre2} Transition wavelengths ($\lambda$) and electric quadrupole transition properties such as line strengths ($S_{vk}$), 
transition probabilities ($A_{vk}$) and oscillator strengths ($f_{kv}$) along with their uncertainties between transitions involving an upper
energy level (denoted by $v$) and lower energy level (denoted by $k$) of Sr$^+$. Values given in square brackets represent the power of 10.}\\
\hline
\hline
\multicolumn{2}{c}{Levels} & $\lambda$  & $S_{vk}$ & $A_{vk}$ & $f_{kv}$  \\ 
\cline{1-2}
Upper $(v)$ & Lower $(k)$ & (in {\AA}) & (in a.u.) & (in $\rm{s^{-1}}$) &  \\
\hline
\hline

$4D_{3/2} $ & $ 5S_{1/2}$    &         6870.066   &1.239$\pm$0.009[2]& 2.26$\pm$0.01[0] &3.21$\pm$0.02[-8]  \\
$4D_{5/2} $ & $ 5S_{1/2} $   &         6740.252   &1.88$\pm$0.01[2]& 2.53$\pm$0.02[0] &5.18$\pm$0.04[-8]  \\
$4D_{5/2}  $ & $ 4D_{3/2}$   &       356709.709   &3.57$\pm$0.02[1]& 1.155$\pm$0.007[-9]&3.30$\pm$0.02[-14]   \\
$5P_{3/2} $ & $ 5P_{1/2}$    &       124772.290   &5.47$\pm$0.02[2] &5.07$\pm$0.02[-6]  &2.36$\pm$0.01[-11] \\
$6S_{1/2} $ & $ 4D_{3/2} $   &         3013.806   &3.64$\pm$0.08[1]& 8.19$\pm$0.19[1]& 5.57$\pm$0.13[-8]    \\
\hline
\hline

\end{longtable}
\begin{longtable}{@{\extracolsep\fill}rlllll@{}}
\caption{\label{parametersbae2} Transition wavelengths ($\lambda$) and electric quadrupole transition properties such as line strengths ($S_{vk}$), 
transition probabilities ($A_{vk}$) and oscillator strengths ($f_{kv}$) along with their uncertainties between transitions involving an upper
energy level (denoted by $v$) and lower energy level (denoted by $k$) of Ba$^+$. Values given in square brackets represent the power of 10.}\\
\hline
\hline
\multicolumn{2}{c}{Levels} & $\lambda$  & $S_{vk}$ & $A_{vk}$ & $f_{kv}$  \\ 
\cline{1-2}
Upper $(v)$ & Lower $(k)$ & (in {\AA}) & (in a.u.) & (in $\rm{s^{-1}}$) &  \\
\hline
\hline

$5D_{3/2} $ & $  6S_{1/2}$   &        20517.652   &1.59$\pm$0.03[2] &1.23$\pm$0.02[-2]&1.55$\pm$0.03[-9]     \\
$5D_{5/2} $ & $  6S_{1/2}$   &        17621.744   &2.49$\pm$0.04[2]  &2.74$\pm$0.05[-2]&3.83$\pm$0.07[-9]     \\
$5D_{5/2} $ & $  5D_{3/2}$   &       124850.959   &4.49$\pm$0.08[1]  &2.77$\pm$0.05[-7]&9.70$\pm$0.17[-13]\\
$6P_{3/2} $ & $ 6P_{1/2}$    &       59142.096   &8.03$\pm$0.07[2] & 3.11$\pm$0.03[-4] & 3.25$\pm$0.03[-10]     \\
$7S_{1/2} $ & $  5D_{3/2}$   &         2667.995   &2.25$\pm$0.08[1]  &9.31$\pm$0.34[1]&4.97$\pm$0.19[-8]   \\
\hline
\hline
\end{longtable}

\vspace{1 cm}

\begin{longtable}{@{\extracolsep\fill}rlllllllllll@{}}
\caption{\label{comparmge2}Comparison of transition probabilities $A_{vk}$ (in $\rm{s^{-1}}$) of the E2 forbidden transitions involving upper
level ($v$) and lower level ($k$) in Mg$^+$ from different works. Values given in square brackets represent the power of 10.}\\
\hline
\hline
\multicolumn{2}{c}{{Transition}}  & \multicolumn{5}{c}{{$A_{vk}$}}  \\ 
\cline{1-2}\cline{3-7}
Upper $(v)$ & Lower $(k)$ &Present& \cite{celik} & \cite{sanjoy}& \cite{fischer} & NIST~\cite{kramida} \\
\hline
\hline
  3$D_{5/2}$&  3$S_{1/2}$&   5.0971$\pm$0.0008[3]&4.94[3] &5.010[3] & & 8.77[3] \\
  3$D_{3/2}$&  3$S_{1/2}$&   5.098$\pm$0.001[3]&4.94[3] &5.010[3] & \\
  4$D_{5/2}$&  3$S_{1/2}$&   7.191$\pm$0.006[2]&7.01[2] &6.580[2] & & 1.04[3] \\
  4$D_{3/2}$&  3$S_{1/2}$&   7.19$\pm$0.02[2]&7.01[2] &6.590[2] & \\
  3$P_{3/2}$&  3$P_{1/2}$& 2.732$\pm$0.001[-11]&2.99[-11] &3.050[-11] &3.81[-11] \\
\hline
\hline
\end{longtable}

\begin{longtable}{@{\extracolsep\fill}rlllllllll@{}}
\caption{\label{parameterse2} Comparison of the transition probabilities $A_{vk}$ (in $\rm{s^{-1}}$) and absorption oscillator strengths $f_{kv}$ in the Ca$^+$, Sr$^+$ and Ba$^+$ alkaline earth metal ions with different works. Values given in parentheses represent the power of 10.}\\
\hline
\hline
\multicolumn{2}{c}{Transition} &  \multicolumn{2}{c}{{$A_{vk}$}} &  \multicolumn{2}{c}{{$f_{kv}$}} \\ 
\cline{1-2}\cline{3-4}\cline{5-6}
Upper $(v)$ & Lower $(k)$ & Present& Others(Th, Exp) & Present& Others(Th, Exp) \\

\hline
\hline
\multicolumn{6}{c}{Ca$^+$}\\
$3D_{3/2}$&$ 4S_{1/2}$   &  0.838$\pm$0.008(0)&  0.8662$\pm$0.0049(0)~\cite{filippin}  &  1.35$\pm$0.01(-8)& 1.393$\pm${0.007}(-8)~\cite{filippin}\\
            &  & & 1.3(0)~\cite{kramida} & & 2.092(-8)~\cite{kramida} \\
$3D_{5/2}$&$4S_{1/2}$   &   0.860$\pm$0.007(0)& 0.8567$\pm$0.0108(0)~\cite{shao2017}   & 2.05$\pm$0.02(-8)&  2.049$\pm${0.025}(-8)~\cite{shao2017}  \\
             &  & & 0.8892$\pm$0.0049(0)~\cite{filippin} &  & 2.127$\pm${0.012}(-8)~\cite{filippin} \\
\hline
\hline
\end{longtable}

\begin{longtable}{@{\extracolsep\fill}rlllllll@{}}
\caption{\label{staticmg}Static dipole polarizabilities (in a.u.) of the ground and excited states of Mg$^+$ with core correlation contribution 
0.43$\pm$0.02 a.u. from the present work. The valence correlation contributions are given as `Main' and `Tail', whereas the core-valence correlation 
contributions are given under `vc'. Other literature values are also quoted for the comparison. The core-valence correlation contributions 
to the tensor components of the polarizabilities are not quoted here due to their negligible magnitudes.}\\
\hline
\hline
{State}  &    \multicolumn{4}{c}{\textbf{$\alpha_{d,0}^{J}$}}   &  \multicolumn{3}{c}{\textbf{$\alpha_{d,2}^{J}$}}\\
\cline{2-5}\cline{6-8} 
           &  Main     &   Tail      & vc    &  Total     &  Main    &     Tail    &   Total   \\
\hline
\hline           
$3S_{1/2}$   &    34.5$\pm$0.9    &    0.07$\pm$0.03  &  -0.01   &   35$\pm$1\\
&&&& 35.05~\cite{mit2009} \\
$4S_{1/2}$    &    552.7$\pm$0.4  &     0.17$\pm$0.08 &  -0.01   &    553.3$\pm$0.4\\
$5S_{1/2}$    &    3757$\pm$5  &     1.8$\pm$0.9     &  -0.01   &   3759$\pm$5\\
$3P_{1/2}$   &     30.9$\pm$0.5 &    0.2$\pm$0.1    &    $\sim$0       &  31.5$\pm$0.5\\
\hline
\hline

\end{longtable}

\begin{longtable}{@{\extracolsep\fill}rlllllll@{}}
\caption{\label{staticca}Static dipole polarizabilities (in a.u.) of the ground and excited states of Ca$^+$ with core correlation contribution 
3.2$\pm$0.2 a.u. from the present work. The valence correlation contributions are given as `Main' and `Tail', whereas the core-valence correlation 
contributions are given under `vc'. Other literature values are also quoted for the comparison. The core-valence correlation contributions 
to the tensor components of the polarizabilities are not quoted here due to their negligible magnitudes.}\\
\hline
\hline
{State}  &    \multicolumn{4}{c}{\textbf{$\alpha_{d,0}^{J}$}}   &  \multicolumn{3}{c}{\textbf{$\alpha_{d,2}^{J}$}}\\
\cline{2-5}\cline{6-8} 
           &  Main     &   Tail      & vc    &  Total     &  Main    &     Tail    &   Total   \\
\hline
\hline           
$4S_{1/2}$  &  72.8$\pm$0.6  &   0.06$\pm$0.03  & -0.08   &   76.0$\pm$0.6    \\
&&&                                          &   76.1$\pm$0.5~\cite{safronova2011} \\
&&&                                          & 75.28~\cite{tang13} \\
&&&                                          & 73.0$\pm$1.5~\cite{sahoo2009a} \\
&&&                                          & 75.49~\cite{mitroy2008}  \\
\hline
\hline
\end{longtable}

\begin{longtable}{@{\extracolsep\fill}rlllllll@{}}
\caption{\label{staticsr}Static dipole polarizabilities (in a.u.) of the ground and excited states of Sr$^+$ with core correlation contribution 
5.8$\pm$0.3 a.u. from the present work. The valence correlation contributions are given as `Main' and `Tail', whereas the core-valence correlation 
contributions are given under `vc'. Other literature values are also quoted for the comparison. The core-valence correlation contributions 
to the tensor components of the polarizabilities are not quoted here due to their negligible magnitudes.}\\
\hline
\hline
{State}  &    \multicolumn{4}{c}{\textbf{$\alpha_{d,0}^{J}$}}   &  \multicolumn{3}{c}{\textbf{$\alpha_{d,2}^{J}$}}\\
\cline{2-5}\cline{6-8} 
           &  Main     &   Tail      & vc    &  Total     &  Main    &     Tail    &   Total   \\
\hline
\hline           
$5S_{1/2}$&  85.7$\pm$0.9 & 0.02$\pm$0.01 & -0.19  &  91.3$\pm$0.9    &  \\
&&&                                                 &  92.2$\pm$0.7~\cite{safr2010} \\
&&&                                                 &  91.30~\cite{dansha2009} \\
&&&                                                 &  89.88~\cite{mitroy2008PRA}  \\
&&&                                                 &  88.29$\pm$1.00~\cite{sahoo2009b}  \\
\hline
\hline
\end{longtable}

\begin{longtable}{@{\extracolsep\fill}rlllllll@{}}
\caption{\label{staticba}Static dipole polarizabilities (in a.u.) of the ground and excited states of Ba$^+$ with core correlation contribution 
10.6$\pm$0.5 a.u. from the present work. The valence correlation contributions are given as `Main' and `Tail', whereas the core-valence correlation 
contributions are given under `vc'. Other literature values are also quoted for the comparison. The core-valence correlation contributions 
to the tensor components of the polarizabilities are not quoted here due to their negligible magnitudes.}\\
\hline
\hline
{State}  &    \multicolumn{4}{c}{\textbf{$\alpha_{d,0}^{J}$}}   &  \multicolumn{3}{c}{\textbf{$\alpha_{d,2}^{J}$}}\\
\cline{2-5}\cline{6-8} 
           &  Main     &   Tail      & vc    &  Total     &  Main    &     Tail    &   Total   \\
\hline
\hline           
$6S_{1/2}$   &   114$\pm$2   & 0.016$\pm$0.008  &   -0.38    & 124$\pm$2   \\
&&&                                             & 124.26$\pm$1.00~\cite{sahoo2009b} \\
&&&                                             & 123.7$\pm$0.5~\cite{barrett2019} \\
&&&                                             & 124.15~\cite{isk2008} \\
&&&                                             & 123.88$\pm$0.05~\cite{snow2007} \\
\hline
\hline
\end{longtable}

\begin{landscape}
\begin{longtable}{@{\extracolsep\fill}rllllllllll@{}}
\caption{\label{staticquadmg}Static quadrupole polarizabilities (in a.u.) of the ground and excited states of Mg$^+$ with core correlation contribution 
0.52$\pm$0.03 a.u. from the present work. The valence correlation contributions are given as `Main' and `Tail', whereas the core-valence correlation 
contributions are given under `vc'. Other literature values are also quoted for the comparison. The core-valence correlation contributions 
to the tensor components of the polarizabilities are not quoted here due to their negligible magnitudes.}\\
\hline
\hline
{State}& \multicolumn{4}{c}{\textbf{$\alpha_{q,0}^{J}$}} & \multicolumn{3}{c}{\textbf{$\alpha_{q,2}^{J}$}} &  \multicolumn{3}{c}{\textbf{$\alpha_{q,4}^{J}$}}\\
\cline{2-5}\cline{6-8}\cline{9-11} 
           &  Main     &   Tail      & vc    &  Total     &  Main    &     Tail    &   Total &  Main    &     Tail    &   Total   \\
\hline
\hline           
$3S_{1/2}$   &  154.69$\pm$0.02   & 0.8$\pm$0.4 & $\sim$0 &156.0$\pm$0.4\\
&&&              & 156.1~\cite{mitroy2008} \\
&&&              & 156.02$\pm$1.27~\cite{sahoo2007} \\
$3P_{1/2}$   &  72947$\pm$35    & 58$\pm$29 & $\sim$0&73005$\pm$45\\
$3P_{3/2}$  &   -36046$\pm$18   &58$\pm$29 &$\sim$0 &-35987$\pm$34 &36261$\pm$18 & -10$\pm$5 &36251$\pm$18\\
$3D_{3/2}$  &  -7555049$\pm$4363    &797$\pm$399 &$\sim$0 &-7554251$\pm$4381  &-5377735$\pm$3116 & -227$\pm$114 & -5377962$\pm$3118\\
$3D_{5/2}$  &   5019851$\pm$2909   &797$\pm$399 &$\sim$0 &5020648$\pm$2936 & -2549856$\pm$1484& -324$\pm$162 &-2550180$\pm$1493 &-1234286$\pm$712 & 3$\pm$1 &-1234283$\pm$712 \\

\hline
\hline
\end{longtable}
\end{landscape}

\begin{longtable}{@{\extracolsep\fill}rllllllllll@{}}
\caption{\label{staticquadca}Static quadrupole polarizabilities (in a.u.) of the ground and excited states of Ca$^+$ with core correlation 
contribution 6.9$\pm$0.3 a.u. from the present work. The valence correlation contributions are given as `Main' and `Tail', whereas the core-valence 
correlation contributions are given under `vc'. Other literature values are also quoted for the comparison. The core-valence correlation contributions
to the tensor components of the polarizabilities are not quoted here due to their negligible magnitudes.}\\
\hline
\hline
{State}  & \multicolumn{4}{c}{\textbf{$\alpha_{q,0}^{J}$}} & \multicolumn{3}{c}{\textbf{$\alpha_{q,2}^{J}$}}& \multicolumn{3}{c}{\textbf{$\alpha_{q,4}^{J}$}}\\
\cline{2-5}\cline{6-8}\cline{9-11} 
           &  Main     &   Tail      & vc    &  Total     &  Main    &     Tail    &   Total   &  Main    &     Tail    &   Total \\
\hline
\hline           
$4S_{1/2}$  &  852$\pm$4  & 15$\pm$8 &$\sim$0  &  874$\pm$9 \\
&&&                & 871$\pm$4~\cite{safronova2011} \\
&&&                & 875.1~\cite{mitroy2008} \\
&&&                & 875$\pm$10~\cite{sukhjit2018} \\
$4P_{1/2}$  & 74854$\pm$381   & 66$\pm$33   &$\sim$0 & 74927$\pm$382 \\          
$4P_{3/2}$  & -35910$\pm$191  & 66$\pm$33 & $\sim$0 &-35837$\pm$194  & 36703$\pm$191 & -14$\pm$7 &36689$\pm$191 \\
$3D_{3/2}$  &  5122$\pm$55 & 66$\pm$33 &$\sim$0 &5195$\pm$64  &    3788$\pm$39  & -18$\pm$9 & 3770$\pm$40\\
&&&                                     & 5225$\pm$103~\cite{sukhjit2018} &&& 3797$\pm$71~\cite{sukhjit2018}  \\
$3D_{5/2}$ & -3494$\pm$36 & 66$\pm$33 &$\sim$0 & -3421$\pm$49 & 1895$\pm$19 & -26$\pm$13 &1869$\pm$23  & 826$\pm$9 &0.2$\pm$0.1 & 826$\pm$9 \\
&&&                                  & -3450$\pm$72~\cite{sukhjit2018} &&& 1883$\pm$36~\cite{sukhjit2018} &&& 831
$\pm$16~\cite{sukhjit2018} \\ 
\hline
\hline
\end{longtable}

\begin{landscape}
\begin{longtable}{@{\extracolsep\fill}rllllllllll@{}}
\caption{\label{staticquadsr}Static quadrupole polarizabilities (in a.u.) of the ground and excited states of Sr$^+$ with core correlation 
contribution 17.1$\pm$0.9 a.u. from the present work. The valence correlation contributions are given as `Main' and `Tail', whereas the core-valence 
correlation contributions are given under `vc'. Other literature values are also quoted for the comparison. The core-valence correlation contributions 
to the tensor components of the polarizabilities are not quoted here due to their negligible magnitudes.}\\
\hline
\hline
{State}  & \multicolumn{4}{c}{\textbf{$\alpha_{q,0}^{J}$}} &\multicolumn{3}{c}{\textbf{$\alpha_{q,2}^{J}$}}&\multicolumn{3}{c}{\textbf{$\alpha_{q,4}^{J}$}}\\
\cline{2-5}\cline{6-8}\cline{9-11}  
           &  Main     &   Tail      & vc    &  Total     &  Main    &     Tail    &   Total &  Main    &     Tail    &   Total  \\
\hline
\hline           
$5S_{1/2}$&  1328$\pm$5 & 30$\pm$15 & $\sim$0 & 1375$\pm$16 \\
&&&                            & 1370$\pm$28\cite{safr2010} \\
&&&                            & 1379$\pm$17~\cite{sukhjit2018} \\
&&&                            & 1346\cite{mitroy2008PRA} \\
&&&                            & 1356.300$\pm$0.315~\cite{jiang2016} \\
\hline
\hline
\end{longtable}
\end{landscape}

\begin{longtable}{@{\extracolsep\fill}rllllllllll@{}}
\caption{\label{staticquadba}Static quadrupole polarizabilities (in a.u.) of the ground and excited states of Ba$^+$ with core correlation 
contribution 46$\pm$2 a.u. from the present work. The valence correlation contributions are given as `Main' and `Tail', whereas the core-valence 
correlation contributions are given under `vc'. Other literature values are also quoted for the comparison. The core-valence correlation contributions 
to the tensor components of the polarizabilities are not quoted here due to their negligible magnitudes.}\\ 
\hline
\hline
{State}  &  \multicolumn{4}{c}{\textbf{$\alpha_{q,0}^{J}$}}  &  \multicolumn{3}{c}{\textbf{$\alpha_{q,2}^{J}$}}&  \multicolumn{3}{c}{\textbf{$\alpha_{q,4}^{J}$}}\\
\cline{2-5}\cline{6-8}\cline{9-11} 
       &  Main  & Tail   & vc &  Total     &  Main &  Tail & Total  &  Main &  Tail & Total \\
\hline
\hline           
$6S_{1/2}$ &  4106$\pm$42 &40$\pm$20 & $\sim$0 &4192$\pm$47        \\
&&&                                           & 4182$\pm$34~\cite{isk2008} \\
&&&                                           & 4270$\pm$27~\cite{sahoo2012} \\
&&&                                           & 4251$\pm$61~\cite{sukhjit2018}  \\ 
&&&                                           & 4091.5~\cite{uisaf2010} \\
\hline
\hline
\end{longtable}

\begin{longtable}{@{\extracolsep\fill}rlllll@{}}
\caption{\label{magicmg}Magic wavelengths $\lambda_{\rm{magic}}$ (in nm) and corresponding polarizability values (in a.u.) for the $3S-3P_{3/2,1/2}$ 
and $3S-3D_{5/2,3/2}$ transitions of Mg$^+$ using the linearly polarized light. The resonant wavelengths (in nm) for the transitions contributing to the 
$nP_{J}$ and $nD_J$ ac polarizabilities are listed as $\lambda_{\rm{res}}$.}\\

\hline
\hline
Resonances & $\lambda_{\rm{res}}$  & $|M_J|$ & $\lambda_{\rm{magic}}$  & $\alpha_d^J(\lambda_{\rm{magic}})$  \\
\hline
\hline
Transition & \multicolumn{4}{c}{\underline{$3S-3P_{1/2}$}}\\ 
		       &  & 1/2 & 276.1$\pm$0.7 & $-1266$ \\
$3P_{1/2}$-$3D_{3/2}$ & 279.160 \\
$3P_{1/2}$-$3S_{1/2}$ & 280.353 \\
$3P_{1/2}$-$4S_{1/2}$ & 292.948 \\
                    && 1/2 & 307$\pm$2  & 202 \\
\end{longtable}

\begin{longtable}{@{\extracolsep\fill}rlllll@{}}
\caption{\label{magicca}Magic wavelengths $\lambda_{\rm{magic}}$ (in nm) and corresponding polarizability values (in a.u.) for the $4S-4P_{3/2,1/2}$ 
and $4S-3D_{5/2,3/2}$ transitions of Ca$^+$ using the linearly polarized light. The resonant wavelengths (in nm) for the transitions contributing to the 
$nP_{J}$ and $nD_J$ ac polarizabilities are listed as $\lambda_{\rm{res}}$.} \\
\hline
\hline
Resonances & $\lambda_{\rm{res}}$ & $|M_J|$ & $\lambda_{\rm{magic}}$ &$\alpha_d^J(\lambda_{\rm{magic}})$  \\
\hline
\hline

Transition & \multicolumn{5}{c}{ \underline{$4S-4P_{1/2}$}} \\

$4P_{1/2}-4D_{3/2}$ & 315.978    \\
            & & 1/2  & 368.08$\pm$0.04 &  $-485$     \\
            & &     &  368.0149 &  $-477.2554$  & \cite{tang13} \\
            & &     &  368.0221$\pm$0.1412~\cite{jiang2017} &\\        
$4P_{1/2}-5S_{1/2}$ & 370.707\\
\end{longtable}

\begin{longtable}{@{\extracolsep\fill}rlllll@{}}
\caption{\label{magicsr}Magic wavelengths $\lambda_{\rm{magic}}$ (in nm) and corresponding polarizability values (in a.u.) for the $5S-5P_{3/2,1/2}$ 
and $5S-4D_{5/2,3/2}$ transitions of Sr$^+$ using the linearly polarized light. The resonant wavelengths (in nm) for the transitions contributing to the 
$nP_{J}$ and $nD_J$ ac polarizabilities are listed as $\lambda_{\rm{res}}$.}\\
\hline
\hline
Resonances & $\lambda_{\rm{res}}$ & $|M_J|$ & $\lambda_{\rm{magic}}$  &  $\alpha_d^J(\lambda_{\rm{magic}})$  \\
\hline
\hline
Transition & \multicolumn{4}{c}{\underline{ $5S-5P_{1/2}$}} \\
$5P_{1/2}-5D_{3/2}$ & 338.167 \\
$5P_{1/2}-6S_{1/2}$ & 416.296 \\
$5P_{1/2}-5S_{1/2}$ & 421.670 \\ 
		    & & 1/2 & 767$\pm$2 & 127   \\
		    & &  & 764.238~\cite{jiang2016} & \\
\end{longtable}

\begin{longtable}{@{\extracolsep\fill}rlllll@{}}
\caption{\label{magicba}Magic wavelengths $\lambda_{\rm{magic}}$ (in nm) and corresponding polarizability values (in a.u.) for the $6S-6P_{3/2,1/2}$ 
and $6S-5D_{5/2,3/2}$ transitions of Ba$^+$ using the linearly polarized light. The resonant wavelengths (in nm) for the transitions contributing to the 
$nP_{J}$ and $nD_J$ ac polarizabilities are listed as $\lambda_{\rm{res}}$.}\\
\hline
\hline
Resonances    &$\lambda_{\rm{res}}$ & $|M_{J}|$ & $\lambda_{\rm{magic}}$  & $\alpha_d^J(\lambda_{\rm{magic}})$ \\
\hline
\hline
Transition & \multicolumn{4}{c}{\underline{$6S-6P_{1/2}$}} \\
$6P_{1/2}-8S_{1/2}$ & 264.805 \\
			 && 1/2 & 323$\pm$2 &  $-93$ \\ 
$6P_{1/2}-6D_{3/2}$ & 389.288\\  
             && 1/2 & 451.72$\pm$0.04 &  $-4557$\\
$6P_{1/2}-7S_{1/2}$ & 452.619   \\
\end{longtable}

\begin{landscape}
\begin{longtable}{@{\extracolsep\fill}rllllllllllll@{}}
\caption{\label{tuneout} The tune-out wavelengths ($\lambda_{\rm{T}}$) in nm for Mg$^+$, Ca$^+$, Sr$^+$, Ba$^+$ using the linearly polarized light. }\\
\hline
\hline
\multicolumn{3}{c}{Mg$^+$} & \multicolumn{3}{c}{Ca$^+$} & \multicolumn{3}{c}{Sr$^+$} & \multicolumn{3}{c}{Ba$^+$}\\
\cline{1-3}\cline{4-6}\cline{7-9}\cline{10-12}
State & $M_{J}$ & $\lambda_{\rm{T}}$ & State & $M_{J}$ & $\lambda_{\rm{T}}$ & State & $M_{J}$ & $\lambda_{\rm{T}}$ & State & $M_{J}$ & $\lambda_{\rm{T}}$ \\
\hline

$3S_{1/2}$  & $1/2$ & 280.110$\pm$0.009  & $4S_{1/2}$  & $1/2$ & 395.80$\pm$0.02  & $5S_{1/2}$  & $1/2$ & 417.04$\pm$0.06  & $6S_{1/2}$ & $1/2$ &480.63$\pm$0.24\\
$3P_{1/2}$  & $1/2$ & 280.95$\pm$0.04  & $4P_{1/2}$  & $1/2$ & 364.13$\pm$0.07  & $5P_{1/2}$  & $1/2$ & 411.94$\pm$0.09  & $6P_{1/2}$ & $1/2$ &444.06$\pm$0.17 \\
            & $1/2$ & 290.05$\pm$0.06  &             & $1/2$ & 431.60$\pm$0.85  &             & $1/2$ & 447.86$\pm$0.83  &            & $1/2$ &526.29$\pm$0.80 \\
            &       &              &             &       &                  &             &       &                  &            & $1/2$ &1071$\pm$6 \\
$3P_{3/2}$  & $1/2$ & 279.29$\pm$0.02  & $4P_{3/2}$  & $1/2$ & {318.08$\pm$0.01}  & $5P_{3/2}$  & $1/2$ & 347.57$\pm$0.01 & $6P_{3/2}$ & $1/2$ &416.65$\pm$0.01 \\
\hline
\hline
\end{longtable}
\end{landscape}

\section{Conclusion}

By applying the relativistic all-order method with single and double excitations and also incorporating the partial triple excitations, the E2 matrix 
elements are evaluated for a large number of transitions between many excited states of the singly charged magnesium, calcium, strontium and barium 
alkaline earth-metal ions. This includes 114, 114, 130 and 96 transitions in Mg$^+$, Ca$^+$, Sr$^+$ and Ba$^+$ respectively and electric quadrupole transition properties are also calculated for these transitions. In addition, the results
of the static as well as dynamic dipole and quadrupole polarizabilities of the ground and low-lying metastable states of the above ions are enlisted.
The magic and tune-out wavelengths inferred by plotting the dynamic polarizabilities these state of the considered alkaline earth-metal ions are also 
tabulated. We have also quoted the estimated uncertainties to our calculations, and compare them with the available literature data. Many of the 
reported quadrupole transition properties associated with the high-lying excited states are presented for the first time. These data will be useful 
for both the atomic spectroscopy analyses and interpreting astrophysical observations. The listed magic and tune-out wavelengths will of immense 
interest for suitably designing ionic traps to conduct high-precision experiments.
\ack

The work of B.A. is supported by DST-SERB(India) Grant No. EMR/2016/001228. The employed all order method was developed in the group of Professor M. S. Safronova of the University of Delaware, USA.

\clearpage

\end{document}